\documentclass[journal, twocolumn, twoside]{IEEEtran}
\usepackage[utf8]{inputenc}
\usepackage{amsmath,amsfonts,amssymb}
\usepackage{booktabs}
\usepackage{geometry}
\usepackage{microtype}
\usepackage{float}
\usepackage{xcolor}
\usepackage{tcolorbox}
\usepackage{listings}
\usepackage{subcaption}
\usepackage{multirow}
\usepackage{cite}
\usepackage{enumitem}
\usepackage{dblfloatfix}

\renewcommand\thesubfigure{\alph{subfigure}}
\definecolor{mGreen}{rgb}{0,0.6,0}
\definecolor{mGray}{rgb}{0.5,0.5,0.5}
\definecolor{mPurple}{rgb}{0.58,0,0.82}
\definecolor{backgroundColour}{rgb}{0.95,0.95,0.92}

\lstdefinestyle{myMatlabStyle}{
    backgroundcolor=\color{backgroundColour},   
    commentstyle=\color{mGreen},
    keywordstyle=\color{magenta},
    numberstyle=\tiny\color{mGray},
    stringstyle=\color{mPurple},
    basicstyle=\ttfamily\footnotesize,
    breakatwhitespace=false,         
    breaklines=true,                 
    captionpos=b,                    
    keepspaces=true,                 
    numbers=left,                    
    numbersep=5pt,                  
    showspaces=false,                
    showstringspaces=false,
    showtabs=false,                  
    tabsize=2,
    language=Matlab
}
\usepackage{tikz}
\usepackage{amsmath,amssymb}
\usepackage{subcaption}
\usetikzlibrary{arrows.meta, positioning, calc, shapes.geometric, shapes.misc}

\title{Timing Skew Estimation of Time-Interleaved ADC based on Inter-Channel Correlations without Computing Derivative of Input Signal}
\author{Bibhu Datta Sahoo and Gabriele Manganaro \vspace{-2em}
\thanks{~Bibhu~Datta~Sahoo is with the Department
of Electrical Engineering, University at Buffalo, Buffalo,
NY, 14260 USA (e-mail: bibhu@buffalo.edu).}% <-this % stops a space
\thanks{Gabriele Manganaro is with Strategic Technology Exploration Platform, Mediatek, Woburn, MA, 01801, USA (e-mail: gabriele.manganaro@ieee.org).}
}

\begin{document}

\maketitle

\begin{abstract}
Time-interleaved analog-to-digital converters (TI-ADCs) enable multi-gigasample-per-second throughput in high-speed applications, but sub-picosecond timing skews among interleaved channels introduce large spectral artifacts that degrade dynamic performance. To continuously track parametric drifts without interrupting normal operation, background calibration is required. However, extracting timing skews under unknown inputs is challenging, and many existing correlation-based methods rely on resource-heavy derivative filters that restrict input bandwidth. This article proposes a novel background timing-skew estimation technique for two-channel TI-ADCs that overcomes these limitations. Closed-form timing-skew expressions are derived using: (a) adjacent-channel cross-correlation combined with single-channel auto-correlation for sinusoidal inputs, and (b) adjacent-channel cross-correlation for Wide-Sense Stationary (WSS) inputs, specifically bandpass-filtered wideband noise and Orthogonal Frequency Division Multiplexed (OFDM) signals. The derived analytical models are validated through comprehensive simulations, offering an efficient framework for real-time background calibration in high-speed data converters.
\end{abstract}
\begin{IEEEkeywords}
time-interleaved analog-to-digital converter (ADC), offset, gain, timing-skew, auto-correlation, cross-correlation, OFDM, wide-sense-stationary (WSS), . \vspace{-1.5em}
\end{IEEEkeywords}

\IEEEpeerreviewmaketitle

\section{Introduction}
Time-interleaved analog-to-digital converters (TI-ADCs) \cite{Black1980} are key for achieving multi-gigasample-per-second (GS/s) throughput in high-speed infrastructure, such as coherent optical communications, 5G/6G millimeter-wave transceivers, ultra-wideband radar, and high-end test instrumentation. This is done by distributing sampling across an array of parallel ADCs. But, semiconductor manufacturing variations, clock path asymmetries, and thermal gradients introduce picosecond-level timing skew, $\Delta t$, among the ADC sampling clocks, resulting in non-uniform sampling across interleaved channels. Although careful design and layout techniques can minimize the timing skew, the residual skew is often still large enough to impact the processing of GHz signals. Because the resulting sampling error voltage for an input signal, $x(t)=A \sin(2\pi f_{in}t)$, scales linearly with the input frequency ($\Delta V \approx 2\pi f_{\text{in}} A \cdot \Delta t$), for sampled signals in the GHz range, even sub-picosecond timing mismatches generate large interleaving artifacts in the digital output spectrum, degrading the Spurious-Free Dynamic Range (SFDR) and Signal-to-Noise-and-Distortion Ratio (SNDR) to levels that are inadequate to many applications \cite{Razavi2013,manganaro2022introduction}. Closed form expressions of the impact of timing-skew on SNDR and SFDR have been proposed in \cite{gustavsson2000cmos,jenq1988digital,petraglia1991analysis,ghosh2021closed}. In higher-order modulation systems like 64-QAM or 256-QAM and multi-carrier OFDM channels, this spectral distortion induces inter-symbol interference (ISI) and sharply degrades Adjacent Channel Leakage Ratio (ACLR) and Bit Error Rates (BER) \cite{Aaron2016,luu2018design}. Consequently, continuous timing-skew estimation and adaptive correction are required to track parametric drifts, including those due to silicon aging, in real time. 

\indent  The timing skew calibration process is a two step process, {\em viz.}, (a) timing-skew estimation and (b) timing-skew correction, which can be done either in foreground, where normal ADC operation is interrupted or in background, where timing skew variation due to supply voltage, temperature, and device ageing can be tracked. As background calibration tracks the environmental variation, it is preferred over a foreground method. But, the timing-skew extraction in the background becomes challenging with unknown inputs. 

\indent Typical background timing-skew estimation techniques  \cite{Razavi2013,Huang2007,Vogel2008,Divi2009,Oshima2009,McNeill2009,Satarzadeh2010,Harris2011,Centurelli2012,Matsuno2013,LeDortz2014,wei20148,LeDuc2016,Straayer2016,le2017fully} used cross-correlations between adjacent ADC channels and cross-correlations between ADC output channels and their derivatives to extract the timing-skew errors. As mentioned in \cite{le2017fully} computing the derivative of a channel for input signals close to Nyquist requires a large finite-impulse response (FIR) filter, thus restricting the skew-estimation technique to signals well below Nyquist. Unfortunately, since the interleaving artifacts' power introduced by timing skews grow with the input signal frequency, the derivative computation becomes computationally challenging precisely where it is most important. An excellent overview of various timing-skew calibration techniques in TI-ADCs is given in  \cite{gu2024timing}. This article proposes a new background timing-skew estimation technique for TI-ADCs suitable for various input signal types. The main contributions of this work are closed-form timing-skew expressions for two-channel ADCs based on: (a) adjacent-channel cross-correlation and auto-correlation for sinusoidal inputs, and (b) adjacent-channel cross-correlation for two types of Wide-Sense Stationary (WSS) signals-namely, bandpass-filtered wideband noise and Orthogonal Frequency Division Multiplexed (OFDM) signals.  

\indent  The paper is organized as follows: Section \ref{sec:overview} provides an overview of the TI-ADC architecture and its non-idealities. Section \ref{sec:correlation} derives a closed-form expression for the adjacent-channel cross-correlations in an $M$-channel TI-ADC with a sinusoidal input. Section \ref{sec:time_skew_sine} and Section \ref{sec:time_skew_wss} derive closed-form expressions for the two-channel timing skew under sinusoidal and wide-sense stationary (WSS) inputs, respectively, and validate these expressions through simulation. Finally, Section \ref{sec:conclusion} concludes the paper. \vspace{-1em}

\begin{figure}[htbp]
\centering

% =========================================================
% SUBFIGURE (a): Block Diagram
% =========================================================
\begin{subfigure}{\linewidth}
\centering
\scalebox{0.78}{
\begin{tikzpicture}[>=Stealth, line width=1pt]

  % --- Analog Input x(t) ---
  \node (in_text) at (-5.6, 0.4) {$x(t)$};
  \draw (in_text.east) -- (-4.2, 0.4); 
  \filldraw[fill=white, draw=black, line width=1pt] (in_text.east) circle (2pt);

  % Junction dots on input bus
  \fill (-4.2, 1.4) circle (2pt);
  \fill (-4.2, 0.4) circle (2pt);
  \fill (-4.2, -0.6) circle (2pt);

  % Bus vertical lines
  \draw (-4.2, 3.0) -- (-4.2, 2.2);
  \draw[line width=1pt] (-4.2, 2.2) -- (-4.2, 1.4);
  \draw[line width=1pt] (-4.2, 1.4) -- (-4.2, 0.4);
  \draw[line width=1pt] (-4.2, 0.4) -- (-4.2, -0.6);
  \draw[line width=1pt] (-4.2, -0.6) -- (-4.2, -1.8);
  \draw (-4.2, -1.8) -- (-4.2, -2.6);

  % Switches before ADCs
  % Channel 0
  \draw (-4.2, 3.0) -- (-3.4, 3.0);
  \fill (-3.4, 3.0) circle (1.8pt);
  \draw (-3.4, 3.0) -- (-2.8, 3.25); % open switch arm
  \fill (-2.6, 3.0) circle (1.8pt);
  \draw[->] (-2.6, 3.0) -- (-1.1, 3.0) node[midway, above=-1pt, font=\normalsize] {$y_0[n]$};
  % Switch Clock Arrow (Thick)
  \draw[->, line width=1.8pt] (-3.1, 3.6) node[above] {$\Phi_0$} -- (-3.1, 3.1);

  % Channel 1
  \draw (-4.2, 1.4) -- (-3.4, 1.4);
  \fill (-3.4, 1.4) circle (1.8pt);
  \draw (-3.4, 1.4) -- (-2.8, 1.65);
  \fill (-2.6, 1.4) circle (1.8pt);
  \draw[->] (-2.6, 1.4) -- (-1.1, 1.4) node[midway, above=-1pt, font=\normalsize] {$y_1[n]$};
  % Switch Clock Arrow (Thick)
  \draw[->, line width=1.8pt] (-3.1, 2) node[above] {$\Phi_1$} -- (-3.1, 1.5);

  % Channel l
  \draw (-4.2, -0.6) -- (-3.4, -0.6);
  \fill (-3.4, -0.6) circle (1.8pt);
  \draw (-3.4, -0.6) -- (-2.8, -0.35);
  \fill (-2.6, -0.6) circle (1.8pt);
  \draw[->] (-2.6, -0.6) -- (-1.1, -0.6) node[midway, above=-1pt, font=\normalsize] {$y_k[n]$};
  % Switch Clock Arrow (Thick)
  \draw[->, line width=1.8pt] (-3.1, 0) node[above] {$\Phi_k$} -- (-3.1, -0.5);

  % Channel M-1
  \draw (-4.2, -2.6) -- (-3.4, -2.6);
  \fill (-3.4, -2.6) circle (1.8pt);
  \draw (-3.4, -2.6) -- (-2.8, -2.35);
  \fill (-2.6, -2.6) circle (1.8pt);
  \draw[->] (-2.6, -2.6) -- (-1.25, -2.6) node[midway, above=-1pt, font=\normalsize] {$y_{M-1}[n]$};
  % Switch Clock Arrow (Thick)
  \draw[->, line width=1.8pt] (-3.1, -2) node[above] {$\Phi_{M-1}$} -- (-3.1, -2.5);

  % --- ADC Blocks (Filled Cyan) ---
  \node[draw, rectangle, fill=cyan!15, minimum width=1.5cm, minimum height=0.75cm, font=\normalsize] (adc0) at (-0.35, 3.0) {$\mathbf{ADC_0}$};
  \node[draw, rectangle, fill=cyan!15, minimum width=1.5cm, minimum height=0.75cm, font=\normalsize] (adc1) at (-0.35, 1.4) {$\mathbf{ADC_1}$};
  \node[draw, rectangle, fill=cyan!15, minimum width=1.5cm, minimum height=0.75cm, font=\normalsize] (adcl) at (-0.35, -0.6) {$\mathbf{ADC_k}$};
  \node[draw, rectangle, fill=cyan!15, minimum width=1.5cm, minimum height=0.75cm, font=\normalsize] (adcm) at (-0.35, -2.6) {$\mathbf{ADC_{M-1}}$};

  % --- Clock Inputs (Thick lines top of each ADC) ---
  \draw[->, line width=1.8pt] (-0.35, 3.8) -- (adc0.north) node[above=7pt, font=\normalsize] {$\Phi_0$};
  \draw[->, line width=1.8pt] (-0.35, 2.2) -- (adc1.north) node[above=7pt, font=\normalsize] {$\Phi_1$};
  \draw[->, line width=1.8pt] (-0.35, 0.2) -- (adcl.north) node[above=7pt, font=\normalsize] {$\Phi_k$};
  \draw[->, line width=1.8pt] (-0.35, -1.8) -- (adcm.north) node[above=7pt, font=\normalsize] {$\Phi_{M-1}$};

  % --- Trapezoidal MUX (Filled Orange) ---
  \node[draw, 
        regular polygon rotate=90, 
        minimum width=6.2cm, minimum height=1.2cm, 
        trapezium, trapezium angle=75, trapezium stretches=true,
        fill=orange!20, font=\large\bfseries, shape border rotate=270] (mux) at (2.8, 0.2) {\rotatebox{-90}{MUX}};

  % Connections from ADCs to MUX
  \draw[->] (adc0.east) -- (mux.north west |- adc0.east) node[midway, above=-1pt, font=\normalsize] {$D_0$};
  \draw[->] (adc1.east) -- (mux.north west |- adc1.east) node[midway, above=-1pt, font=\normalsize] {$D_1$};
  \draw[->] (adcl.east) -- (mux.north west |- adcl.east) node[midway, above=-1pt, font=\normalsize] {$D_k$};
  \draw[->] (adcm.east) -- (mux.north west |- adcm.east) node[midway, above=-1pt, font=\normalsize] {$D_{M-1}$};

  % Vertical dotted line on MUX input bus
  \draw[dotted, line width=0.9pt] (1.3, -1.1) -- (1.3, -2.1);

  % --- Digital Output D_out ---
  \draw[->] (mux.east) -- (4.2, 0.2) 
    node[midway, above=0pt, font=\normalsize] {$D_{out}$} 
    node[midway, below=-1pt, font=\normalsize] {$f_s$};
\end{tikzpicture}
}
\caption{}
\end{subfigure}
% =========================================================
% SUBFIGURE (b): Timing Diagram
% =========================================================
\begin{subfigure}{\linewidth}
\centering
\scalebox{0.78}{
\begin{tikzpicture}[>=Stealth, line width=0.8pt]

  % --- Y-Axis Channel Labels ---
  \node[anchor=east, font=\normalsize] at (-0.2, 0) {$\Phi_0$};
  \node[anchor=east, font=\normalsize] at (-0.2, -1.1) {$\Phi_1$};
  \draw[dotted, line width=0.9pt] (-0.5, -1.5) -- (-0.5, -1.9);
  \node[anchor=east, font=\normalsize] at (-0.2, -2.3) {$\Phi_k$};
  \draw[dotted, line width=0.9pt] (-0.5, -2.7) -- (-0.5, -3.1);
  \node[anchor=east, font=\normalsize] at (-0.2, -3.5) {$\Phi_{M-1}$};

  % --- Timing Arrows (Top) ---
  \draw[<->] (0.5, 0.7) -- (1.5, 0.7) node[midway, above, font=\normalsize] {$T_s$};
  \draw[<->] (1.5, 0.7) -- (7.3, 0.7) node[midway, above, font=\normalsize] {$(M-1)T_s$};

  % --- Waveform ADC_0 ---
  \draw (0,0) -- (0.5,0) -- (0.5,0.35) -- (0.8,0.35) -- (0.8,0) -- (2.5,0);
  \draw[dotted, line width=0.9pt] (2.5,0) -- (3.2,0);
  \draw (3.2,0) -- (4.8,0);
  \draw[dotted, line width=0.9pt] (4.8,0) -- (5.5,0);
  \draw (5.5,0) -- (7.3,0) -- (7.3,0.35) -- (7.6,0.35) -- (7.6,0) -- (8.5,0);

  % --- Waveform ADC_1 ---
  \draw (0,-1.1) -- (1.5,-1.1) -- (1.5,-0.75) -- (1.8,-0.75) -- (1.8,-1.1) -- (2.5,-1.1);
  \draw[dotted, line width=0.9pt] (2.5,-1.1) -- (3.2,-1.1);
  \draw (3.2,-1.1) -- (4.8,-1.1);
  \draw[dotted, line width=0.9pt] (4.8,-1.1) -- (5.5,-1.1);
  \draw (5.5,-1.1) -- (8.5,-1.1);

  % --- Waveform ADC_l ---
  \draw (0,-2.3) -- (2.5,-2.3);
  \draw[dotted, line width=0.9pt] (2.5,-2.3) -- (3.2,-2.3);
  \draw (3.2,-2.3) -- (4.0,-2.3) -- (4.0,-1.95) -- (4.3,-1.95) -- (4.3,-2.3) -- (4.8,-2.3);
  \draw[dotted, line width=0.9pt] (4.8,-2.3) -- (5.5,-2.3);
  \draw (5.5,-2.3) -- (8.5,-2.3);

  % --- Waveform ADC_{M-1} ---
  \draw (0,-3.5) -- (2.5,-3.5);
  \draw[dotted, line width=0.9pt] (2.5,-3.5) -- (3.2,-3.5);
  \draw (3.2,-3.5) -- (4.8,-3.5);
  \draw[dotted, line width=0.9pt] (4.8,-3.5) -- (5.5,-3.5);
  \draw (5.5,-3.5) -- (6.3,-3.5) -- (6.3,-3.15) -- (6.6,-3.15) -- (6.6,-3.5) -- (8.5,-3.5);

\end{tikzpicture}
}
\caption{}
\end{subfigure}

\caption{(a) Time-interleaved ADC with $M$ channels and (b) its timing relationship.}
\label{fig:ti_adc_full}
\end{figure}
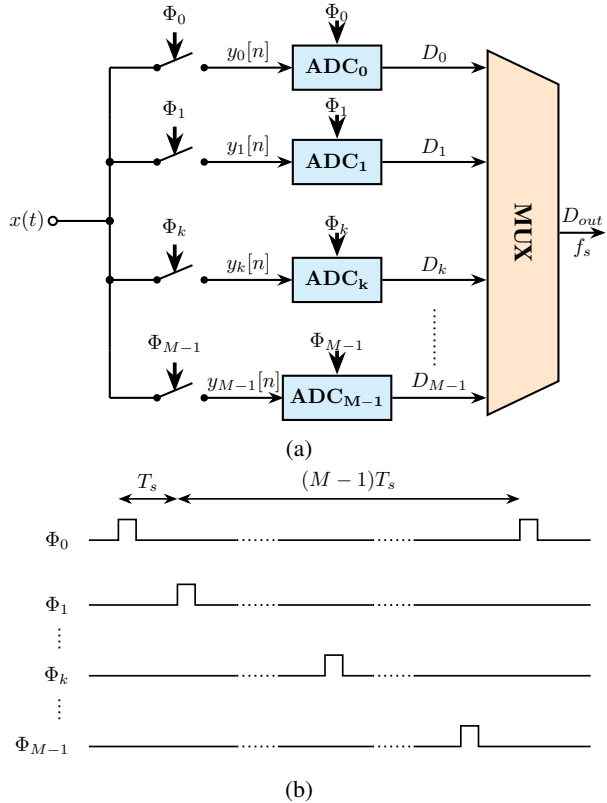

 \vspace{-1.5em}

\section{Overview of Time-Interleaved ADC}
\label{sec:overview}
\indent Figure \ref{fig:ti_adc_full}(a) shows the block diagram of an $M$-channel TI-ADC, where $x(t)$ is the input signal, $y_i[n]$, for $i=0$ to $(M-1)$ are the sampled signals being digitized by $\mathbf{ADC_i}$, for $i=0$ to $(M-1)$ at a sample-rate of $f_s/M$ to create digital outputs $D_i[n]$, for $i=0$ to $(M-1)$. The digital outputs of each channel are demultiplexed using the block $\mathbf{MUX}$ to create the digital representation, $D_{out}$, of the input signal, $x(t)$, at a rate of $f_s$. Figure \ref{fig:ti_adc_full}(b) shows the clock waveforms $\Phi_i$ for $i=0$ to $(M-1)$ that clock each ADC. Circuit non-idealities result in each of the channels having different offset, gain, input bandwidth, non-uniformly spaced clocks $\Phi_k$. Thus, the sampling switch and the ADC in each channel in Fig. \ref{fig:ti_adc_full}(a), shown also in Fig. \ref{fig:adc_models}(a), can be modeled as shown in Fig. \ref{fig:adc_models}(b), where, $g_k$, $o_k$, $\Delta t_k$ are the gain, offset, and timing-skew, respectively. It is worth noting that the amount by which the clock is skewed is modeled by delaying the input signal $x(t)$ by that same amount.  These channel mismatches introduce distinct dynamic errors, often referred to as interleaving spectral artifacts, that degrade system performance by: (a) adding fixed-frequency deterministic spurs at integer multiples of $f_s/M$ regardless of the input signal (offset mismatch); (b) amplitude-modulating the input signal to create image spurs at $\pm f_{in} + k(f_s/M)$ (gain mismatch); and (c) phase-modulating the input signal due to timing skew, which generates image spurs at the same frequencies ($\pm f_{in} + k(f_s/M)$), but with amplitudes that scale proportionally with input frequency. Since the primary focus of this work is timing-skew estimation, the remainder of this article considers timing-skew non-idealities exclusively.\\
 \begin{figure}[htbp]
\centering

% =========================================================
% SUBFIGURE (a): Ideal ADC
% =========================================================
\begin{subfigure}{\linewidth}
\centering
\resizebox{0.65\linewidth}{!}{%
\begin{tikzpicture}[
  >=Stealth,
  thick,
  block/.style={draw, rectangle, thick, minimum height=1.1cm, minimum width=1.2cm, align=center, font=\large},
  adc/.style={draw, rectangle, fill=cyan!15, line width=1.3pt, minimum height=1.1cm, minimum width=1.5cm, align=center, font=\large},
  circlep/.style={draw, circle, thick, inner sep=0pt, minimum size=0.65cm, font=\large},
  dot/.style={circle, fill=black, inner sep=1.2pt}
]

  % Nodes & Connections
  \node at (1.2, 0) (x_a) {$x(t)$};
  
  % Switch
  \draw (x_a) -- (2.1, 0);
  \filldraw[fill=white, draw=black, line width=1.5pt]  (x_a.east) circle (2pt);
  \node[dot] (sw1_a) at (2.1, 0) {};
  \draw[thick] (2.1, 0) -- (2.7, 0.3);
  \node[dot] (sw2_a) at (3.0, 0) {};

  % Switch Clock Arrow
  \draw[->] (2.4, 0.5) node[above] {$\Phi_k$} -- (2.4, 0.2);
  
  % Line to ADC & ADC block
  \draw[->] (sw2_a) -- (4.5, 0) node[midway, above=2pt] {$y_k[n]$};
  \node[adc] (adc_a) at (5.3, 0) {$\mathbf{ADC}_k$};
  
  % Clock input
  \draw[<-] (adc_a.north) -- ++(0, 0.6) node[above] {$\Phi_k$};
  
  % Output
  \draw[->] (adc_a.east) -- ++(1.4, 0) node[midway, above=1pt] {$D_k$};
  
  % Bounding Dashed Box
  \draw[dashed, line width=1pt] (1.8, -0.7) rectangle (7.1, 1.7);
  \node[anchor=north west, font=\sffamily\bfseries\Large] at (1.85, 1.65) {Ideal ADC};

\end{tikzpicture}%
}
\caption{}
%\label{fig:ideal_adc}
\end{subfigure}

% =========================================================
% SUBFIGURE (b): Non-ideal ADC
% =========================================================
\begin{subfigure}{\linewidth}
\centering
\resizebox{1.02\linewidth}{!}{%
\begin{tikzpicture}[
  >=Stealth,
  thick,
  block/.style={draw, rectangle, thick, minimum height=1.1cm, minimum width=1.2cm, align=center, font=\large},
  adc/.style={draw, rectangle, fill=cyan!15, line width=1.3pt, minimum height=1.1cm, minimum width=1.5cm, align=center, font=\large},
  circlep/.style={draw, circle, thick, inner sep=0pt, minimum size=0.65cm, font=\large},
  dot/.style={circle, fill=black, inner sep=1.2pt}
]

  % Input
  \node[font=\large] at (-0.6, -3.8) (x_b) {$x(t)$};
  
  % Time Shift Block
  \node[block] (ts) at (1.2, -3.8) {time\\shift\\$\Delta t_k$};
  \draw[->] (x_b) -- (ts.west);
  \filldraw[fill=white, draw=black, line width=2pt]  (x_b.east) circle (2pt);
  
  % Multiplier
  \node[circlep] (mult) at (3.8, -3.8) {$\times$};
  \draw[->] (ts.east) -- (mult.west) node[midway, above=2pt] {$x(t - \Delta t_k)$};
  
  % Gain Input g_l
  \draw[->] (3.8, -5.0) node[below, font=\large] {$g_{_k}$} -- (mult.south);
  
  % Adder
  \node[circlep] (add) at (5.3, -3.8) {$+$};
  \draw[->] (mult.east) -- (add.west);
  
  % Offset Input o_l
  \draw[->] (5.3, -5.0) node[below, font=\large] {$o_{_k}$} -- (add.south);
  
  % Switch
  \draw (add.east) -- (6.3, -3.8);
  \node[dot] (sw1_b) at (6.3, -3.8) {};
  \draw[thick] (6.3, -3.8) -- (6.8, -3.5);
  \node[dot] (sw1_b) at (7.0, -3.8) {};
  \draw (7.0, -3.8) -- (7.5, -3.8);
  
  % Switch Clock Arrow
  \draw[->] (6.65, -2.8) node[above, font=\large] {$\Phi_k$} -- (6.65, -3.5);
  
  % ADC Block
  \node[adc, font=\large] (adc_b) at (8.3, -3.8) {$\mathbf{ADC}_k$};
  \draw (7.0, -3.8) -- (adc_b.west);
  
  % Clock input phi_0
  \draw[<-] (adc_b.north) -- ++(0, 0.6) node[above, font=\large] {$\Phi_k$};
  
  % Output D_0
  \draw[->] (adc_b.east) -- ++(1.4, 0) node[midway, xshift=-4pt, above=1pt,font=\large] {$D_k$};
  
  % Bounding Dashed Box
  \draw[dashed, line width=1pt] (0, -5.8) rectangle (10.1, -2.0);
  \node[anchor=north west, font=\sffamily\bfseries\Large] at (0.05, -2.05) {Non-ideal ADC};

  \node[font=\large] at (8.1, -5.5) {$\Phi_k=(k M  + l) \cdot T_s$};
\end{tikzpicture}%
}
\caption{}
%\label{fig:nonideal_adc}
\end{subfigure}

\caption{$k^{th}$-Channel ADC model under (a) ideal condition and (b) non-ideal condition.}
\label{fig:adc_models}
\end{figure}
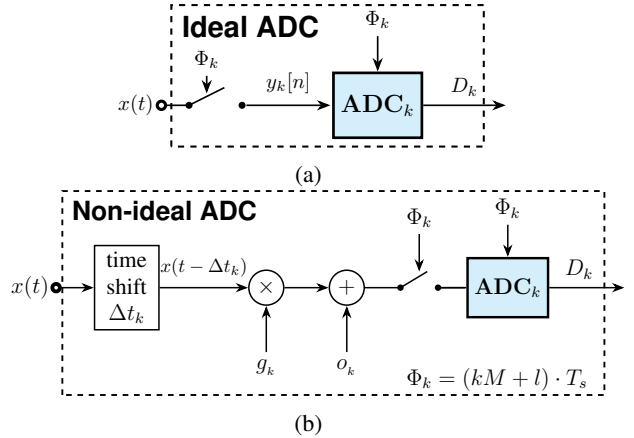
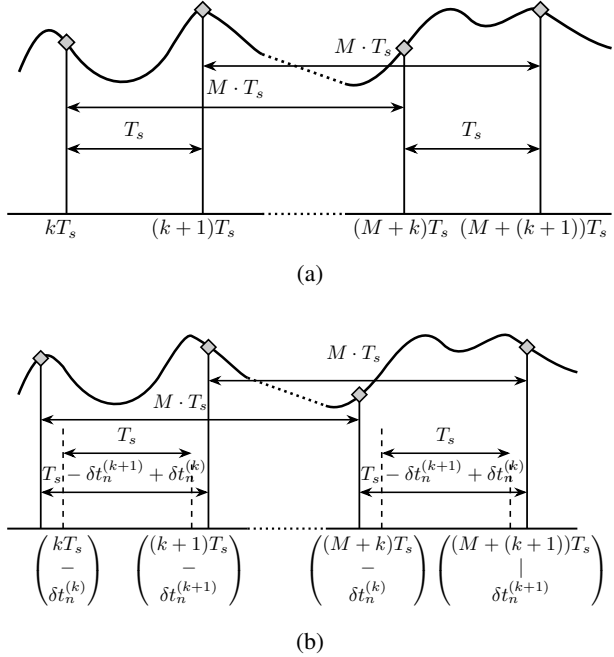
\begin{figure}[htbp]
\centering

% =========================================================
% SUBFIGURE (a): Ideal Sampling
% =========================================================
\begin{subfigure}{\linewidth}
\centering
\resizebox{\linewidth}{!}{%
\begin{tikzpicture}[>=Stealth, line width=0.8pt]

  % --- Baseline ---
  \draw[line width=1pt] (0,0) -- (4.3,0);
  \draw[dotted, line width=1.1pt] (4.3,0) -- (5.7,0);
  \draw[line width=1pt] (5.7,0) -- (10.2,0);

  % --- Analog Waveform ---
  \draw[line width=1.3pt] (0.2,2.4) .. controls (0.6,3.4) and (0.8,3.1) .. (1.0,2.9) 
                          .. controls (1.5,2.1) and (2.2,2.0) .. (2.7,2.7) 
                          .. controls (3.0,3.2) and (3.2,3.5) .. (3.3,3.45) 
                          .. controls (3.8,3.2) and (4.1,2.8) .. (4.3,2.7);
                          
  \draw[dotted, line width=1.3pt] (4.3,2.7) -- (5.7,2.2);
  
  \draw[line width=1.3pt] (5.7,2.2) .. controls (6.0,2.1) and (6.3,2.3) .. (6.7,2.8) 
                          .. controls (7.2,3.5) and (7.5,3.6) .. (7.8,3.3) 
                          .. controls (8.2,2.9) and (8.7,3.6) .. (9.0,3.45) 
                          .. controls (9.4,3.2) and (9.8,2.9) .. (10.2,2.8);

  % --- Sampling Vertical Lines & Points ---
  % Point 1: kT_s
  \draw[line width=0.9pt] (1.0,0) -- (1.0,2.9);
  \node[diamond, draw, fill=black!20, minimum size=1mm, inner sep=2pt] at (1.0,2.9) {};

  % Point 2: (k+1)T_s
  \draw[line width=0.9pt] (3.3,0) -- (3.3,3.45);
  \node[diamond, draw, fill=black!20, minimum size=1mm, inner sep=2pt] at (3.3,3.45) {};

  % Point 3: MT_s + kT_s
  \draw[line width=0.9pt] (6.7,0) -- (6.7,2.8);
  \node[diamond, draw, fill=black!20, minimum size=1mm, inner sep=2pt] at (6.7,2.8) {};

  % Point 4: MT_s + (k+1)T_s
  \draw[line width=0.9pt] (9.0,0) -- (9.0,3.45);
  \node[diamond, draw, fill=black!20, minimum size=1mm, inner sep=2pt] at (9.0,3.45) {};

  % --- Timing Interval Arrows ---
  % Level 1: T_s
  \draw[<->, line width=0.8pt] (1.0, 1.1) -- (3.3, 1.1) node[midway, above=1pt, font=\normalsize] {$T_s$};
  \draw[<->, line width=0.8pt] (6.7, 1.1) -- (9.0, 1.1) node[midway, above=1pt, font=\normalsize] {$T_s$};

  % Level 2: M * T_s (Left to Middle)
  \draw[<->, line width=0.8pt] (1.0, 1.8) -- (6.7, 1.8) node[midway, above=1pt, font=\normalsize] {$M \cdot T_s$};

  % Level 3: M * T_s (Middle to Right)
  \draw[<->, line width=0.8pt] (3.3, 2.5) -- (9.0, 2.5) node[midway, above=1pt, xshift=-4pt, font=\normalsize] {$M \cdot T_s$};

  % --- Axis Labels (Rotated 90 Degrees) ---
  \node[rotate=0, anchor=west, font=\normalsize] at (0.5, -0.25) {$k T_s$};
  \node[rotate=0, anchor=west, font=\normalsize] at (2.3, -0.25) {$(k+1) T_s$};
  \node[rotate=0, anchor=west, font=\normalsize] at (5.7, -0.25) {$(M + k) T_s$};
  \node[rotate=0, anchor=west, font=\normalsize] at (7.5, -0.25) {$(M + (k+1)) T_s$};

\end{tikzpicture}%
}
\caption{}
%\label{fig:ideal_sampling}
\end{subfigure}

\vspace{0.5cm}

% =========================================================
% SUBFIGURE (b): Non-Uniform Sampling with Timing Offset
% =========================================================
\begin{subfigure}{\linewidth}
\centering
\resizebox{\linewidth}{!}{%
\begin{tikzpicture}[>=Stealth, line width=0.8pt]

  % --- Baseline ---
  \draw[line width=1pt] (0,0) -- (4.3,0);
  \draw[dotted, line width=1.1pt] (4.3,0) -- (5.7,0);
  \draw[line width=1pt] (5.7,0) -- (10.2,0);

  % --- Analog Waveform ---
  \draw[line width=1.3pt] (0.2,2.4) .. controls (0.6,3.4) and (0.8,3.1) .. (1.0,2.9) 
                          .. controls (1.5,2.1) and (2.2,2.0) .. (2.7,2.7) 
                          .. controls (3.0,3.2) and (3.2,3.5) .. (3.3,3.45) 
                          .. controls (3.8,3.2) and (4.1,2.8) .. (4.3,2.7);
                          
  \draw[dotted, line width=1.3pt] (4.3,2.7) -- (5.7,2.2);
  
  \draw[line width=1.3pt] (5.7,2.2) .. controls (6.0,2.1) and (6.3,2.3) .. (6.7,2.8) 
                          .. controls (7.2,3.5) and (7.5,3.6) .. (7.8,3.3) 
                          .. controls (8.2,2.9) and (8.7,3.6) .. (9.0,3.45) 
                          .. controls (9.4,3.2) and (9.8,2.9) .. (10.2,2.8);

  % --- Ideal Reference Lines (Dashed) ---
  \draw[dashed, line width=0.8pt] (1.0,0) -- (1.0,1.8);
  \draw[dashed, line width=0.8pt] (3.3,0) -- (3.3,1.15);
  \draw[dashed, line width=0.8pt] (6.7,0) -- (6.7,1.8);
  \draw[dashed, line width=0.8pt] (9.0,0) -- (9.0,1.15);

  % --- Actual Sampling Lines (Solid) & Points ---
  % Point 1: kT_s - dt_n^{(k)}
  \draw[line width=0.9pt] (0.6,0) -- (0.6,3.1);
  \node[diamond, draw, fill=black!20, minimum size=1mm, inner sep=2pt] at (0.6,3.05) {};

  % Point 2: (k+1)T_s - dt_n^{(k+1)}
  \draw[line width=0.9pt] (3.6,0) -- (3.6,3.3);
  \node[diamond, draw, fill=black!20, minimum size=1mm, inner sep=2pt] at (3.6,3.25) {};

  % Point 3: MT_s + kT_s - dt_n^{(k)}
  \draw[line width=0.9pt] (6.3,0) -- (6.3,2.5);
  \node[diamond, draw, fill=black!20, minimum size=1mm, inner sep=2pt] at (6.3,2.4) {};

  % Point 4: MT_s + (k+1)T_s - dt_n^{(k+1)}
  \draw[line width=0.9pt] (9.3,0) -- (9.3,3.3);
  \node[diamond, draw, fill=black!20, minimum size=1mm, inner sep=2pt] at (9.3,3.25) {};

  % --- Timing Interval Arrows ---
  % Level 1: T_s - dt_n^{(k+1)} + dt_n^{(k)}
  \draw[<->, line width=0.8pt] (0.6, 0.65) -- (3.6, 0.65) node[midway, above=1pt, font=\normalsize] {$T_s - \delta t_n^{(k+1)} + \delta t_n^{(k)}$};
  \draw[<->, line width=0.8pt] (6.3, 0.65) -- (9.3, 0.65) node[midway, above=1pt, font=\normalsize] {$T_s - \delta t_n^{(k+1)} + \delta t_n^{(k)}$};

  % Level 2: T_s (Ideal)
  \draw[<->, line width=0.8pt] (1.0, 1.35) -- (3.3, 1.35) node[midway, above=1pt, font=\normalsize] {$T_s$};
  \draw[<->, line width=0.8pt] (6.7, 1.35) -- (9.0, 1.35) node[midway, above=1pt, font=\normalsize] {$T_s$};

  % Level 3: M * T_s (Left to Middle)
  \draw[<->, line width=0.8pt] (0.6, 1.95) -- (6.3, 1.95) node[midway, above=1pt, xshift=-10pt,  font=\normalsize] {$M \cdot T_s$};

  % Level 4: M * T_s (Middle to Right)
  \draw[<->, line width=0.8pt] (3.6, 2.65) -- (9.3, 2.65) node[midway, above=1pt, xshift=-7pt, yshift=5pt,  font=\normalsize] {$M \cdot T_s$};

  % --- Axis Labels (Rotated 90 Degrees) ---
  \node[rotate=0, anchor=west, font=\normalsize] at (0.3, -0.75) {$\begin{pmatrix} k T_s \\ - \\ \delta t_n^{(k)} \end{pmatrix}$}; 

  \node[rotate=0, anchor=west, font=\normalsize] at (2.1, -0.75) {$\begin{pmatrix} (k+1) T_s \\ - \\ \delta t_n^{(k+1)} \end{pmatrix}$}; 

  \node[rotate=0, anchor=west, font=\normalsize] at (5.2, -0.75) {$\begin{pmatrix} (M + k) T_s \\ - \\ \delta t_n^{(k)} \end{pmatrix}$}; 

  \node[rotate=0, anchor=west, font=\normalsize] at (7.5, -0.75) {$\begin{pmatrix} (M + (k+1)) T_s \\ | \\ \delta t_n^{(k+1)} \end{pmatrix}$};

\end{tikzpicture}%
}
\caption{}
%\label{fig:nonuniform_sampling}
\end{subfigure}

\caption{Time-interleaved sampling relationship under (a) ideal condition and (b) timing mismatch.} \vspace{-1em}
\label{fig:ti_sampling_mismatch}
\end{figure}
\indent The ideal sampling instant of channel $k$ is given by \eqref{eq:ideal_samp_edges} and is shown in Fig. \ref{fig:ti_sampling_mismatch}(a). 
\begin{equation}
t_{n}^{(k)} = nMT_{s} + kT_{s}.
\label{eq:ideal_samp_edges}
\end{equation}
The sampling instant with timing-mismatch is given by \eqref{eq:real_samp_edges} and is shown in Fig. \ref{fig:ti_sampling_mismatch}(b).
\begin{equation}
t_{n}^{(k)} = nMT_{s} + kT_{s} + \delta t_{k}.
\label{eq:real_samp_edges}
\end{equation}
For the subsequent analysis we will assume $\delta t_0=0$ for simplicity. 

\section{Adjacent-Channel Cross-Correlation in an $M$-channel TI-ADC for Sinusoidal Input}
\label{sec:correlation}
Consider a sinusoidal input signal, $x(t)=Acos(\omega_0 t+\theta)$, being sampled by an $M$ TI-ADC with timing skew, $\delta t_k$ for $k=0$ to $(M-1)$. The sampled signal for $k^{th}$-channel is given by \eqref{eq:ykn}.
\begin{equation}
    y_k[n]=A\cos(\omega_0 t_n^{(k)} + \theta)
    \label{eq:ykn}
\end{equation}

Since $\delta t_k$ is very small we can write \eqref{eq:ykn} using Taylor's series expansion with only the first term as given by \eqref{eq:ch_out1},
\begin{equation}
y_k[n] = A \cos(\omega_{0}\tau_{n}^{(k)} + \theta) - \delta t_{k} A \omega_{0} \sin(\omega_{0}\tau_{n}^{(k)} + \theta),
\label{eq:ch_out1}
\end{equation}
where, $\tau_{n}^{(k)} = nMT_{s} + kT_{s}$ (the nominal instant). The second term in \eqref{eq:ch_out1} is the skew induced disturbance. Note that the skew induced disturbance is proportional to the instantaneous slope of the input signal.  

To simplify the analysis we will define two terms $s_k[n]=A \cos(\omega_{0}\tau_{n}^{(k)}+ \theta)$ and $c_k[n]=A \omega_{0} \sin(\omega_{0}\tau_{n}^{(k)} + \theta)$, which results in $y_k[n]$ being given by \eqref{eq:ch_out2}.
\begin{equation}
y_k[n] = s_k[n] - \delta t_{k} \omega_{0} c_k[n],
\label{eq:ch_out2}
\end{equation}

\indent The zero-lag cross correlation \cite{peebles2015probability} between channels $i$ and $j$ is given by \eqref{eq:ch_cr_cor_ij_1}.
\begin{equation}
R_{ij} = \lim_{N\rightarrow\infty} \frac{1}{N} \sum_{n=0}^{N-1} y_{i}[n]y_{j}[n].
\label{eq:ch_cr_cor_ij_1}
\end{equation}
Substituting, \eqref{eq:ch_out2} in \eqref{eq:ch_cr_cor_ij_1} we get \eqref{eq:ch_cr_cor_ij_2}.
\begin{equation}
R_{ij} = \lim_{N\rightarrow\infty} \frac{1}{N} \sum_{n=0}^{N-1} (s_i[n] - \delta t_{i} \omega_{0} c_i[n])(s_j[n] - \delta t_{j} \omega_{0} c_j[n]) \nonumber \\
\label{eq:ch_cr_cor_ij_2}
\end{equation}
The following time-averages, {\em i.e.,} taken over many periods, are useful in obtaining the closed form-expressions for \eqref{eq:ch_cr_cor_ij_2}:
\begin{eqnarray}
     \langle s_is_j \rangle = \frac{A^2}{2}\cos(\omega_0(\tau_n^{(i)}-\tau_n^{(j)}))=\frac{A^2}{2}\cos(\omega_0 \tau_{ij})\\
     \langle c_is_j \rangle = \frac{A^2}{2}\sin(\omega_0(\tau_n^{(i)}-\tau_n^{(j)}))=\frac{A^2}{2}\sin(\omega_0 \tau_{ij})
\end{eqnarray}
where, $\tau_{ij}=\tau_n^{(i)}-\tau_n^{(j)}$. \\
\indent Thus, the zero-lag cross-correlation between channel $i$ and $j$ for sinusoidal input is given by \eqref{eq:cor_sine_in}, where the term corresponding to $\delta t_i \delta t_j$ is ignored.
\begin{equation}
    R_{ij}=\underbrace{\frac{A^2}{2}\cos(\omega_0 \tau_{ij})}_{R_{ij}^{ideal}} +\underbrace{\frac{A^2}{2}\omega_0(\delta t_i - \delta t_j)\sin(\omega_0 \tau_{ij})}_{R_{ij}^{skew}}
    \label{eq:cor_sine_in}
\end{equation}
Note that $\tau_{ij}=T_S$ for two adjacent channels, thus resulting in the following closed form expression for the cross-correlation of signals of any two adjacent channels of an $M$-channel TI-ADC:
\begin{align}
  R_{ij}^{ideal}&=\frac{A^2}{2}\cos(\omega_0 T_S) \\
  R_{ij}^{skew} &= \frac{A^2}{2}\omega_0(\delta t_i - \delta t_j)\sin(\omega_0 T_S)
\end{align}
A compact matrix form notation of the cross-correlation in an $M$-channel TI-ADC can be defined as in \eqref{eq:m_ch_cross},
\begin{equation}
R = \frac{A^{2}}{2}\left[\mathbf{C} - \omega_{0}\mathbf{S} \odot (\overrightarrow{\mathbf{1}}\overrightarrow{\mathbf{\delta t}}^{T} - \overrightarrow{\mathbf{\delta t}}\overrightarrow{\mathbf{1}}^{T})\right]
\label{eq:m_ch_cross}
\end{equation}
where, $\overrightarrow{\mathbf{\delta t}} = [\delta t_{0}, \delta t_{1}, \dots, \delta t_{M-1}]^{T}$, $C_{ij} = \cos(\omega_{0}T_S)$, $S_{ij} = \sin(\omega_{0}T_S)$, $\odot$ denotes element-wise multiplication (Hadamard product), $\overrightarrow{\mathbf{1}}$ is an {\em all-ones} column vector.

\indent In TI-ADCs, the number of sub-ADC channels ($M$) is typically designed as a {\em power-of-two}-such as $M = 2^1 = 2$, $M = 2^2 = 4$, or $M = 2^3 = 8$. This binary structure allows complex, $M$-channel timing skew estimation to be simplified using a hierarchical, {\em divide-and-conquer} approach. Rather than calculating and correcting all channel timing offsets simultaneously, the array is recursively split into sub-pairs and solved iteratively as basic two-channel problems. Thus, in the subsequent sections the timing-skew estimation is derived for $2$-channel case only.
\vspace{-1em}
\section{Two-Channel Timing-Skew Estimation with Sine-Wave Input}
\label{sec:time_skew_sine} 
\begin{figure}[H]
   \centering
   \resizebox{\columnwidth}{!}{%
   \begin{tikzpicture}[
  >=Stealth,
  line width=1.5pt, % Thick lines throughout
  sqnode/.style={draw=black, fill=black!20, rectangle, ultra thick, inner sep=4pt},
  circnode/.style={draw=black, fill=black!20, circle, ultra thick, inner sep=4pt},
  labelstyle/.style={rotate=90, anchor=west, font=\large, fill=white, inner sep=1.5pt}
]

  % --- Waveform Coordinates with TS+Delta (2.2) > TS-Delta (1.0) ---
  \coordinate (P0) at (0.0, 0.3);
  \coordinate (P1) at (0.5, 0.9);   % y0[n]
  \coordinate (P2) at (2.5, 2.2);   % y1[n]    (x1 + 2.0)
  \coordinate (P3) at (3.7, 1.7);   % y0[n+1]  (x2 + 1.2)
  \coordinate (P4) at (5.7, 2.0);   % y1[n+1]  (x3 + 2.0)
  \coordinate (P5) at (6.9, 1.8);   % y0[n+2]  (x4 + 1.2)
  \coordinate (P6) at (8.9, 0.9);   % y1[n+2]  (x5 + 2.0)
  \coordinate (P7) at (10.1, 1.2);  % y0[n+3]  (x6 + 1.2)
  \coordinate (P8) at (12.1, 0.4);  % y1[n+3]  (x7 + 2.0)
  \coordinate (P9) at (13.0, 0.5);

  % --- Vertical Dotted Lines through Sampling Points (Drawn first) ---
  \draw[dotted, line width=1.2pt] (0.5, 2.2) -- (0.5, -1.0);
  \draw[dotted, line width=1.2pt] (2.5, 3.0) -- (2.5, -1.0);
  \draw[dotted, line width=1.2pt] (3.7, 3.0) -- (3.7, -1.0);
  \draw[dotted, line width=1.2pt] (5.7, 3.6) -- (5.7, -1.0);
  \draw[dotted, line width=1.2pt] (6.9, 3.6) -- (6.9, -1.0);
  \draw[dotted, line width=1.2pt] (8.9, 2.7) -- (8.9, -1.0);
  \draw[dotted, line width=1.2pt] (10.1, 3.0) -- (10.1, -1.0);
  \draw[dotted, line width=1.2pt] (12.1, 2.2) -- (12.1, -1.0);

  % --- Smooth Analog Signal Curve ---
  \draw[ultra thick] 
    (P0) .. controls (0.2, 0.6) .. 
    (P1) .. controls (1.3, 1.8) and (1.8, 2.3) .. 
    (P2) .. controls (3.1, 1.8) and (3.3, 1.6) .. 
    (P3) .. controls (4.4, 2.2) and (5.2, 2.2) .. 
    (P4) .. controls (6.3, 1.7) and (6.6, 2.1) .. 
    (P5) .. controls (7.6, 1.4) and (8.3, 0.8) .. 
    (P6) .. controls (9.5, 1.0) and (9.8, 1.5) .. 
    (P7) .. controls (10.8, 0.8) and (11.5, 0.3) .. 
    (P8) .. controls (12.6, 0.3) .. (P9);

  % --- Sampling Nodes ---
  \node[sqnode] at (P1) {};
  \node[circnode] at (P2) {};
  \node[sqnode] at (P3) {};
  \node[circnode] at (P4) {};
  \node[sqnode] at (P5) {};
  \node[circnode] at (P6) {};
  \node[sqnode] at (P7) {};
  \node[circnode] at (P8) {};

  % --- Rotated Labels (Masks dotted lines via fill=white) ---
  \node[labelstyle] at (0.5, 1.25) {$\mathbf{y_0[n]}$};
  \node[labelstyle] at (2.7, 2.55) {$\mathbf{y_1[n]}$};
  \node[labelstyle] at (3.7, 2.05) {$\mathbf{y_0[n+1]}$};
  \node[labelstyle] at (5.9, 2.35) {$\mathbf{y_1[n+1]}$};
  \node[labelstyle] at (6.9, 2.15) {$\mathbf{y_0[n+2]}$};
  \node[labelstyle] at (9.1, 1.25) {$\mathbf{y_1[n+2]}$};
  \node[labelstyle] at (10.1, 1.55) {$\mathbf{y_0[n+3]}$};
  \node[labelstyle] at (12.3, 0.75) {$\mathbf{y_1[n+3]}$};

  % --- Time Annotations Between Dotted Lines ---
  \draw[<->, ultra thick] (0.5, -0.6) -- (2.5, -0.6) 
    node[midway, below=2pt, font=\normalsize] {$\mathbf{T_S + \Delta}$};

  \draw[<->, ultra thick] (2.5, -0.6) -- (3.7, -0.6) 
    node[midway, below=2pt, font=\normalsize] {$\mathbf{T_S - \Delta}$};

  \draw[<->, ultra thick] (3.7, -0.6) -- (5.7, -0.6) 
    node[midway, below=2pt, font=\normalsize] {$\mathbf{T_S + \Delta}$};

  \draw[<->, ultra thick] (5.7, -0.6) -- (6.9, -0.6) 
    node[midway, below=2pt, font=\normalsize] {$\mathbf{T_S - \Delta}$};

  \draw[<->, ultra thick] (6.9, -0.6) -- (8.9, -0.6) 
    node[midway, below=2pt, font=\normalsize] {$\mathbf{T_S + \Delta}$};

  \draw[<->, ultra thick] (8.9, -0.6) -- (10.1, -0.6) 
    node[midway, below=2pt, font=\normalsize] {$\mathbf{T_S - \Delta}$};

  \draw[<->, ultra thick] (10.1, -0.6) -- (12.1, -0.6) 
    node[midway, below=2pt, font=\normalsize] {$\mathbf{T_S + \Delta}$};

\end{tikzpicture}%
      }
   \caption{Non-uniform sampling timing mismatch diagram with adjusted intervals.} \vspace{-1em}
   \label{fig:timing_mismatch_2ch}
\end{figure}
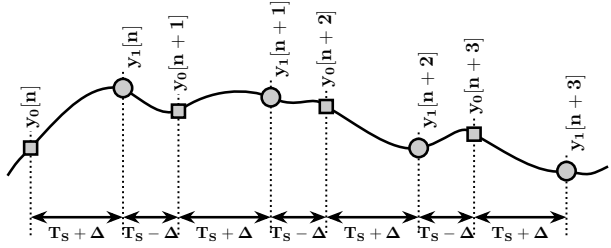

\indent For a 2-channel ADC the timing skew for channel-0 and channel-1 is $\delta t_0=0$ and $\delta t_1 = \Delta$ as shown in Fig. \ref{fig:timing_mismatch_2ch}. Based on \eqref{eq:cor_sine_in}, the cross-correlation between channel-0 and channel-1 is given by \eqref{eq:r01}.
\begin{equation}
    R_{01}=\frac{A^2}{2}\cos(\omega_0 T_S) - \frac{A^2}{2}\omega_0 \Delta \sin(\omega_0T_S).
    \label{eq:r01}
\end{equation}
Similarly, the cross-correlation between channel-1, {\em i.e.,} $y_1[n]$, and the following sample of channel-0, {\em i.e.,} $y_0[n+1]$, can be given by \eqref{eq:r10}.
\begin{equation}
    R_{10}=\frac{A^2}{2}\cos(\omega_0 T_S) + \frac{A^2}{2}\omega_0 \Delta \sin(\omega_0T_S).
    \label{eq:r10}
\end{equation}
From \eqref{eq:r01} and \eqref{eq:r10} the closed form-expression of the timing mismatch in a 2-channel ADC can be obtained as given by \eqref{eq:tskew_exp}.
\begin{equation}
    \Delta = \frac{1}{2}\frac{R_{10}-R_{01}}{\frac{A^2}{2}\omega_0 \sin(\omega_0 T_S)}.
    \label{eq:tskew_exp}
\end{equation}
From \eqref{eq:tskew_exp} it is apparent that the timing-skew can only be obtained if the exact frequency of the sine-wave is known. 

\indent In order to obtain a closed-form expression for the timing-skew, $\Delta$, we rewrite \eqref{eq:r01} and \eqref{eq:r10} as \eqref{eq:r01_mod} and \eqref{eq:r10_mod}, respectively, using the small angle approximation\footnote{For small angle, $\alpha$, $\sin(\alpha)\approx \alpha$ and $\cos(\alpha)=1$.},
\begin{eqnarray}
    R_{01} = \frac{A^2}{2}\cos(\theta_S + \theta_{\Delta}) \label{eq:r01_mod}, \\
    R_{10} = \frac{A^2}{2}\cos(\theta_S - \theta_{\Delta}) \label{eq:r10_mod},
\end{eqnarray}
where, $\theta_S=\omega_0 T_S$ and $\theta_{\Delta}=\omega_0 \Delta$. The auto-correlation of channel-0 is given by \eqref{eq:auto_corr}.
\begin{equation}
    R_{00} = \mathbb{E}[y_0^2[n]] = \frac{A^2}{2}
    \label{eq:auto_corr}
\end{equation}
By normalizing out the amplitude ($A$), we can obtain the normalized correlations as follows shown in \eqref{eq:x1} and \eqref{eq:x2}.
\begin{eqnarray}
    x_1 &=& \frac{R_{01}}{R_{00}} = \cos(T_S\omega_0 + \Delta \omega_0) \label{eq:x1}\\
    x_2 &=& \frac{R_{12}}{R_{00}} = \cos(T_S\omega_0 - \Delta \omega_0) \label{eq:x2}
\end{eqnarray}

Using basic trigonometric identities and small-angle approximation, the sum and difference of \eqref{eq:x1} and \eqref{eq:x2} can be written as \eqref{eq:sum_x} and \eqref{eq:diff_x}, respectively.
\begin{flalign}
    & \frac{x_1 + x_2}{2} = \cos(T_S\omega_0)\cos(\omega_0 \Delta ) = \cos(T_S\omega_0) \hfill \label{eq:sum_x} \\
    & \frac{x_1 - x_2}{2} = -\sin(T_S\omega_0)\sin(\omega_0 \Delta ) = - \omega_0\Delta\sin(T_S\omega_0) \label{eq:diff_x}
\end{flalign}
From \eqref{eq:sum_x} and \eqref{eq:diff_x} $T_S\omega_0$ and $\omega_0\Delta$ can be obtained as shown in \eqref{eq:TsOmega} and \eqref{eq:DeltaOmega}, respectively.
\begin{eqnarray}
    T_S\omega_0 &=& \cos^{-1}\left ( \frac{x_1 + x_2}{2}\right) \label{eq:TsOmega}\\
    \Delta \omega_0 &=& -\frac{\frac{x_1 - x_2}{2}}{\sin(T_S \omega_0)} = -\frac{\frac{x_1 - x_2}{2}}{\sqrt{1-(\frac{x_1 + x_2}{2})^2}} \label{eq:DeltaOmega}
\end{eqnarray}

Taking the ratio of \eqref{eq:TsOmega} and \eqref{eq:DeltaOmega} the closed form expression for relative skew is obtained as shown in \eqref{eq:closed_form_delta}.
%\begin{tcolorbox}[colback=gray!20, colframe=black]
\begin{equation}
    \frac{\Delta}{T_S} = -\frac{x_1 - x_2}{2\sqrt{1-\left(\frac{x_1 + x_2}{2}\right)^2}} \cdot \left[ \cos^{-1}\left ( \frac{x_1 + x_2}{2}\right)  \right]^{-1}.
    \label{eq:closed_form_delta}
\end{equation}
%\end{tcolorbox}
\indent In summary without knowing the frequency $\omega_0$ or the amplitude $A$, the digital signal processor can calculate relative skew $\Delta/T_S$ by only measuring correlations, {\em i.e.}, $R_{00}$, $R_{01}$, and $R_{10}$.

\indent The closed-form expression for the relative timing skew, $\Delta / T_s$ in \eqref{eq:closed_form_delta}, was verified using a two-channel TI-ADC model configured with the following parameters:
\begin{itemize}
    \item \textbf{Sampling Rate:} $f_s = 1\text{ GSPS}$ ($T_s = 1\text{ ns}$)
    \item \textbf{Channel Sampling Rate:} $f_{\text{ch}} = f_s / 2 = 500\text{ MSPS}$
    \item \textbf{Resolution:} 12-bit ideal quantizer
    \item \textbf{Reference Voltage:} $V_{\text{ref}} = 1\text{ V}$ (Full-Scale Range: $\pm 1\text{ V}$)
    \item \textbf{Timing Skew:} Swept from $0.3\%$ to $3\%$ of $T_s$ in increments of $0.3\%$. The absolute skew is computed as:
    \begin{equation}
        \Delta = \frac{\text{skew}_{\%}}{100} \cdot T_s
        \label{eq:per_skew}
    \end{equation}
    \item \textbf{Autocorrelation Averaging Window:} $N = 16\,384 = 2^{14}\text{ samples}$
\end{itemize}

\begin{figure}[htbp]
    \centering
    \includegraphics[scale=0.15]{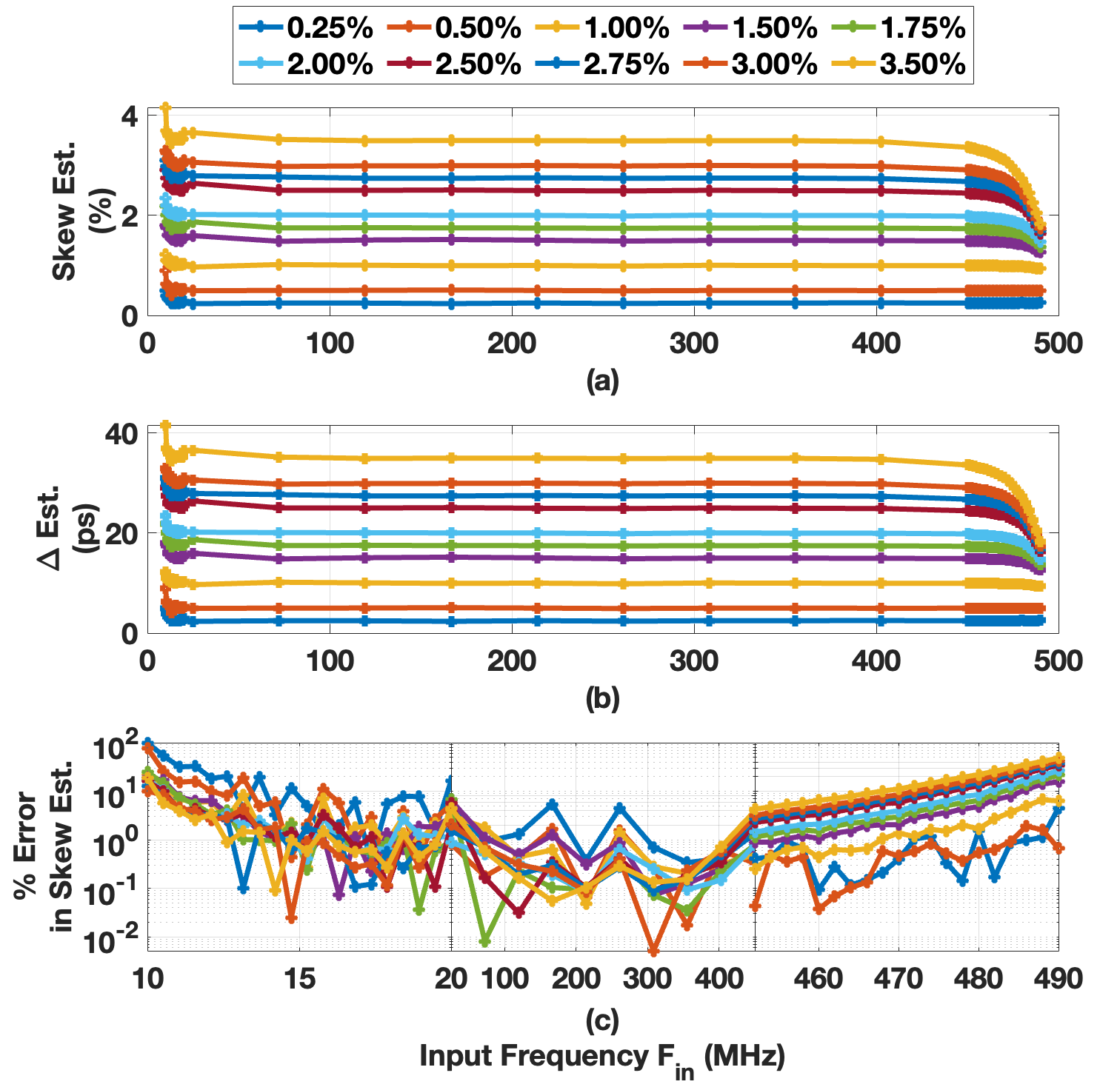}
    \caption{(a) Estimated \% skew, (b) estimated delay $\Delta$ (ps), and (c) \% error in skew estimation. } \vspace{-1em}
    \label{fig:SkewEstimationAnalysis}
\end{figure}

Figures~\ref{fig:SkewEstimationAnalysis}(a) and \ref{fig:SkewEstimationAnalysis}(b) illustrate the estimated skew (in percent) and the absolute timing skew $\Delta$ versus input frequency $f_{\text{in}}$, respectively. The input frequency was swept from $10\text{ MHz}$ to $490\text{ MHz}$ in steps of $53.3\text{ MHz}$. Figure~\ref{fig:SkewEstimationAnalysis}(c) plots the percentage error in skew estimation, defined as:
\begin{align}
\text{\% Err. Skew Est.} &= \left( \frac{|\text{A} - \text{B}|}{\text{B}} \right) \times 100,
\end{align}
where $\text{A} = \text{Estimated Skew (\%)}$ and $\text{B} = \text{Actual Skew (\%)}$. To highlight the substantial variations at low and high frequencies, a logarithmic scale is used for the y-axis for Fig. \ref{fig:SkewEstimationAnalysis}(c). Additionally, the x-axis is split into three distinct segments: $\text{DC}\text{--}0.05\,f_S/2$, $0.05\,f_S/2\text{--}0.8\,f_S/2$, and $>0.8\,f_S/2$.
As observed in Fig.~\ref{fig:SkewEstimationAnalysis}, the estimation error is significant at lower input frequencies. At low frequencies, the magnitude of the timing-skew-induced spurious tone falls below the quantization noise floor of the 12-bit ADC, degrading estimation accuracy. 

Similarly, as $f_{\text{in}}$ approaches the Nyquist frequency ($f_s / 2 = 500\text{ MHz}$), the estimation error rises sharply as the small-angle approximation used to derive \eqref{eq:sum_x} and \eqref{eq:diff_x} do not hold anymore. Consequently, accurate sine-wave-based skew estimation requires the input frequency to be maintained within $10\%$ to $90\%$ of the Nyquist bandwidth:
\begin{equation}
    0.1 \left(\frac{f_s}{2}\right) < f_{\text{in}} < 0.9 \left(\frac{f_s}{2}\right)
\end{equation} 

\section{Two-Channel Timing Skew Estimation with Wide-Sense Stationary (WSS) Input}
\label{sec:time_skew_wss}

\indent For a WSS input signal, the autocorrelation with lag $\tau$ is given by $R_x(\tau) = \mathbb{E}[x(t)x(t+\tau)]$. Thus, for two-channel case with skew $\Delta$, we have $R_{01}$: $\mathbb{E}[y_0[n]y_1[n]] = R_x(T_S + \Delta)$ and $R_{10}$: $\mathbb{E}[y_1[n]y_0[n+1]] = R_x(T_S - \Delta)$. Expanding both terms around the nominal sampling step $T_S$ using a Taylor series we obtain the following:
\begin{eqnarray}
    R_{01} \approx R_x(T_S) + \Delta R'_x(T_S) + \frac{\Delta^2}{2} R''_x(T_S), \\
    R_{10} \approx R_x(T_S) - \Delta R'_x(T_S) + \frac{\Delta^2}{2} R''_x(T_S),
\end{eqnarray}
where, $R'_x(T_S)$ and $R''_x(T_S)$ are the $1^{st}$-derivative and $2^{nd}$-derivative of $R_x(T_S)$ {\em with-respect-to} $\Delta$, respectively. Subtracting $R_{10}$ from $R_{01}$, $\Delta$ can be obtained as shown in \eqref{eq:delta_wss}.
\begin{eqnarray}
    R_{01} - R_{12} &\approx& 2 \Delta R'_x(T_S) \nonumber \\
    \implies \Delta &\approx& \frac{R_{01} - R_{12}}{2 R'_x(T_S)} \label{eq:delta_wss}
\end{eqnarray}
Thus, to compute $\Delta$, we must find the slope of the autocorrelation of lag $T_S$ {\em wrt} $\Delta$. Typically, $R'_x(T_S)$ is obtained using \eqref{eq:auto_deriv}, which requires obtaining derivative of $y_1[n]$, {\em i.e.}, sampled signal in channel-1, which can be a signal close to Nyquist \cite{le2017fully}. 
\begin{equation}
    R_x'(T_s) = \mathbb{E}[y_0[n]y_1'[n]] 
    \label{eq:auto_deriv}
\end{equation}
\subsection{Discrete-time Differentiator}
\begin{figure}[H]
     \centering
    \includegraphics[scale=0.15]{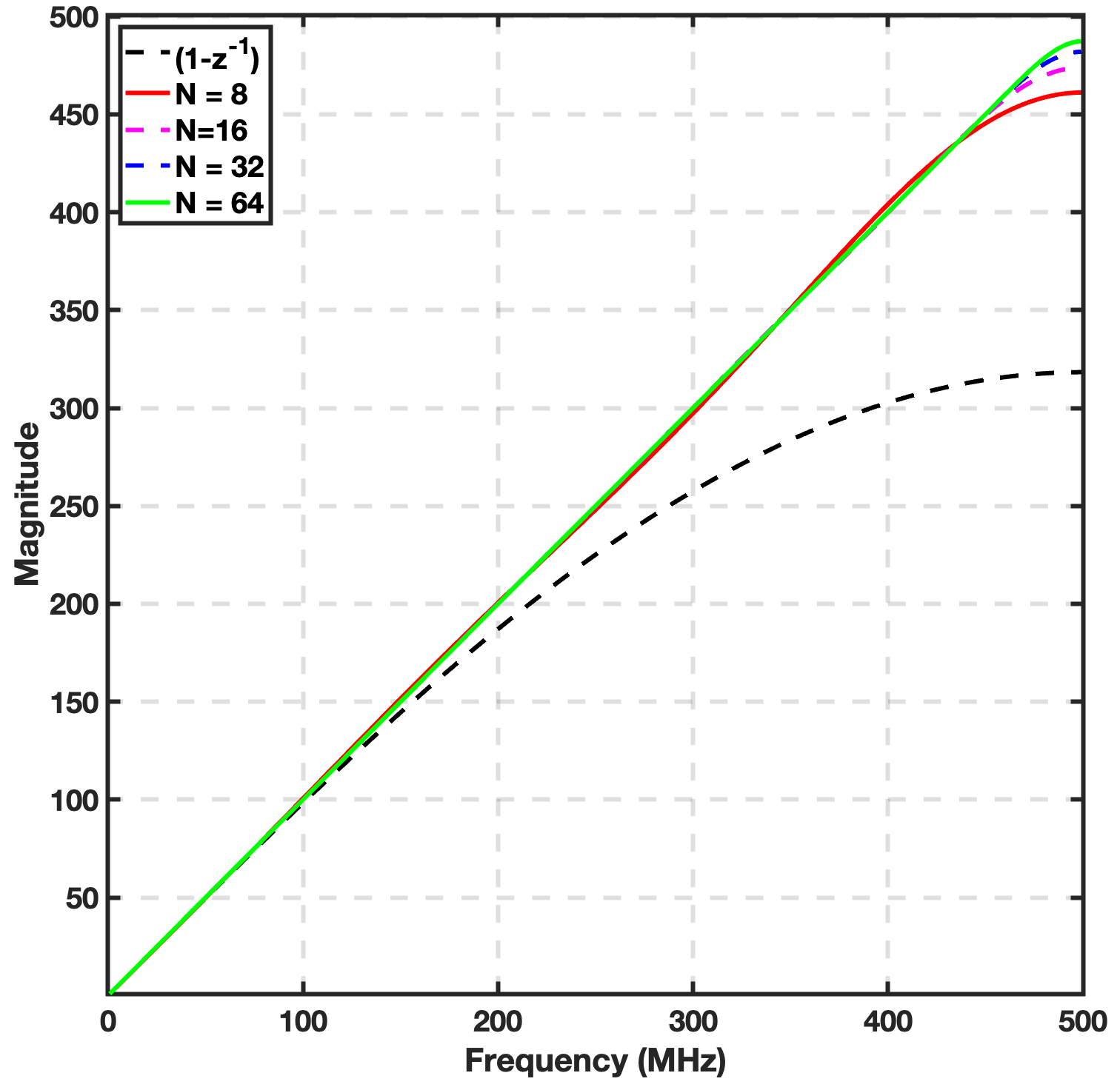} \vspace{-1.5em}
    \caption{Magnitude response of differentiators with different length, $N$.} \vspace{-1em}
    \label{fig:differentiator_response} 
\end{figure}

\begin{figure}[!h]
     \centering 
     \vspace{-1em}
    \includegraphics[scale=0.15]{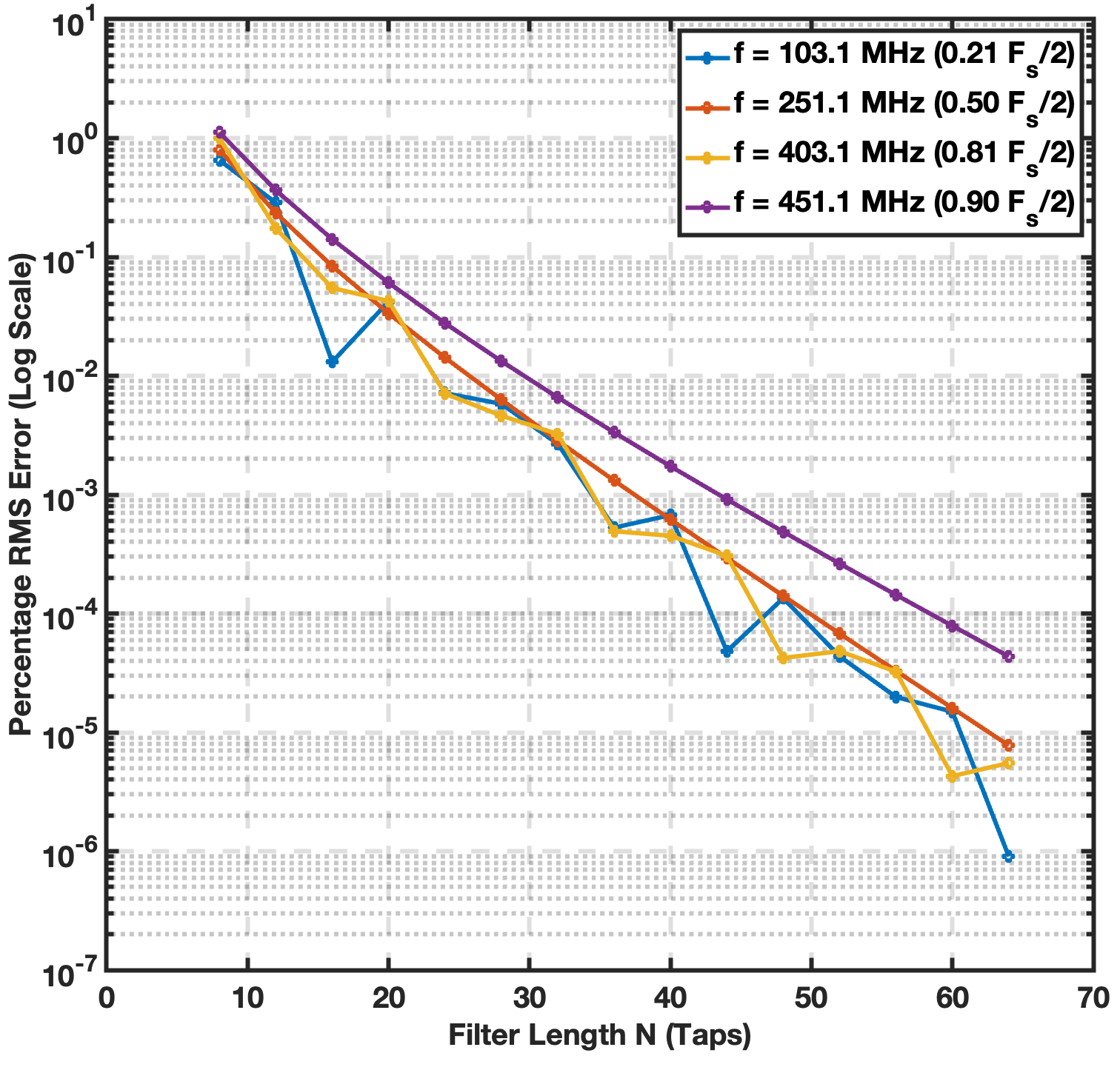} \vspace{-2em}
    \caption{Differentiator Error vs. Filter Length Across Frequencies.} \vspace{-0.5em}
    \label{fig:differentiator_error_vs_freq} 
\end{figure}

\indent Computation of the derivative of a signal which is close to Nyquist is very compute intensive as discussed below. In continuous time, differentiation in the frequency domain corresponds to multiplication by $s$, meaning the ideal frequency response is $H(j\Omega) = j\Omega$. In the discrete-time domain, the ideal frequency response is periodic with period $2\pi$:
\begin{equation}
    H(e^{j\omega}) = j\omega, \quad \text{for } \vert{}\omega\vert{} < \pi
\end{equation}
The simplest discrete-time approximation of a derivative comes from the backward difference formula, $\frac{y[n] - y[n-1]}{T_s}$. Ignoring the scale factor $1/T_s$, the system transfer function and time-domain difference equation is given directly as shown below:
\begin{align}
  H_1(z) = 1 - z^{-1}  \\
  y[n] = x[n] - x[n-1]
\end{align}
For low frequencies ($\omega \approx 0$), $\sin(\omega/2) \approx \omega/2$, giving $H_1(z=e^{j\omega}) \approx j\omega e^{-j\omega/2}$. This accurately mirrors $j\omega$ near $\text{DC}$, but as frequency increases, $\sin(\omega/2)$ deviates significantly from a linear slope \cite{proakis2013digital}.\\
\indent To construct an accurate FIR differentiator across the full Nyquist range, we take the Inverse Discrete-Time Fourier Transform (IDTFT) of the ideal frequency response $H_{\text{ideal}}(e^{j\omega}) = j\omega$ as shown in \eqref{eq:IDFT}, resulting in a non-causal impulse response as given in \eqref{eq:fir_diff}. As \eqref{eq:fir_diff} is non-causal, windowing is used to truncate it to a finite length $N$, resulting in ripples and significant computation error. As the length $N$ of the filter increases the error decreases. Large $N$ results in large number of multiplications needed in the differentiator.  
\begin{equation}
    h_{\text{ideal}}[n] = \frac{1}{2\pi} \int_{-\pi}^{\pi} j\omega \, e^{j\omega n}\, d\omega
    \label{eq:IDFT}
\end{equation}
\begin{equation}
    h_{\text{ideal}}[n] = \begin{cases} 0, & n = 0 \\ \frac{(-1)^n}{n} = \frac{\cos(\pi n)}{n}, & n \neq 0 \end{cases}
    \label{eq:fir_diff}
\end{equation}
However, Parks-McClellan (Remez Exchange) algorithm \cite{proakis2013digital}, can realize differentiators that, relatively speaking, can differentiate signals up to $80\%$ of Nyquist with reasonable filter length $N$ in many practical cases. Figure \ref{fig:differentiator_response} shows the magnitude response of simple differentiator, $H_d(z)=(1-z^{-1})$ and differentiators of various lengths realized using Parks-McClellan method. It is clear from Fig. \ref{fig:differentiator_response}, that a filter length of at least $N=16$ is needed to differentiate signal approaching $80\%$ of Nyquist, though obviously the actual numerical values shown in this figure can be used to determine the suitability and actual parameter range for given system specifications. \vspace{-1em}
Figure \ref{fig:differentiator_error_vs_freq} shows the differentiator error vs. filter length ($N$) across frequencies. It is clear from Fig. \ref{fig:differentiator_error_vs_freq} that to achieve 12-bit accuracy ($<0.025\%$) for signal frequency around $80\%$ of Nyquist $N\approx 28$ length filter is needed \cite{le2017fully}. \\
\indent The $N\approx 28$-tap finite-impulse response (FIR) filter that generates $y_1'[n]$ has to operate at a speed of $f_s/2$ for the $2$-channel TI-ADC and at $f_s/M$ for an $M$-channel TI-ADC, thereby requiring poly-phase based realization of filters \cite{le2017fully,proakis2013digital,mitra2010digital}. Section \ref{sec:skew_est_no_deriv}, describes a method to obtain the skew without computing the derivative of a signal, thereby significantly reducing the hardware complexity.  
\vspace{-1em}
\subsection{Estimation of $\Delta$ without Computing $R_x'(T_S)$ } 
\label{sec:skew_est_no_deriv}
\begin{figure*}
    \centering
    % Row 1
    \begin{subfigure}[b]{0.48\textwidth}
        \centering
        \includegraphics[width=\textwidth]{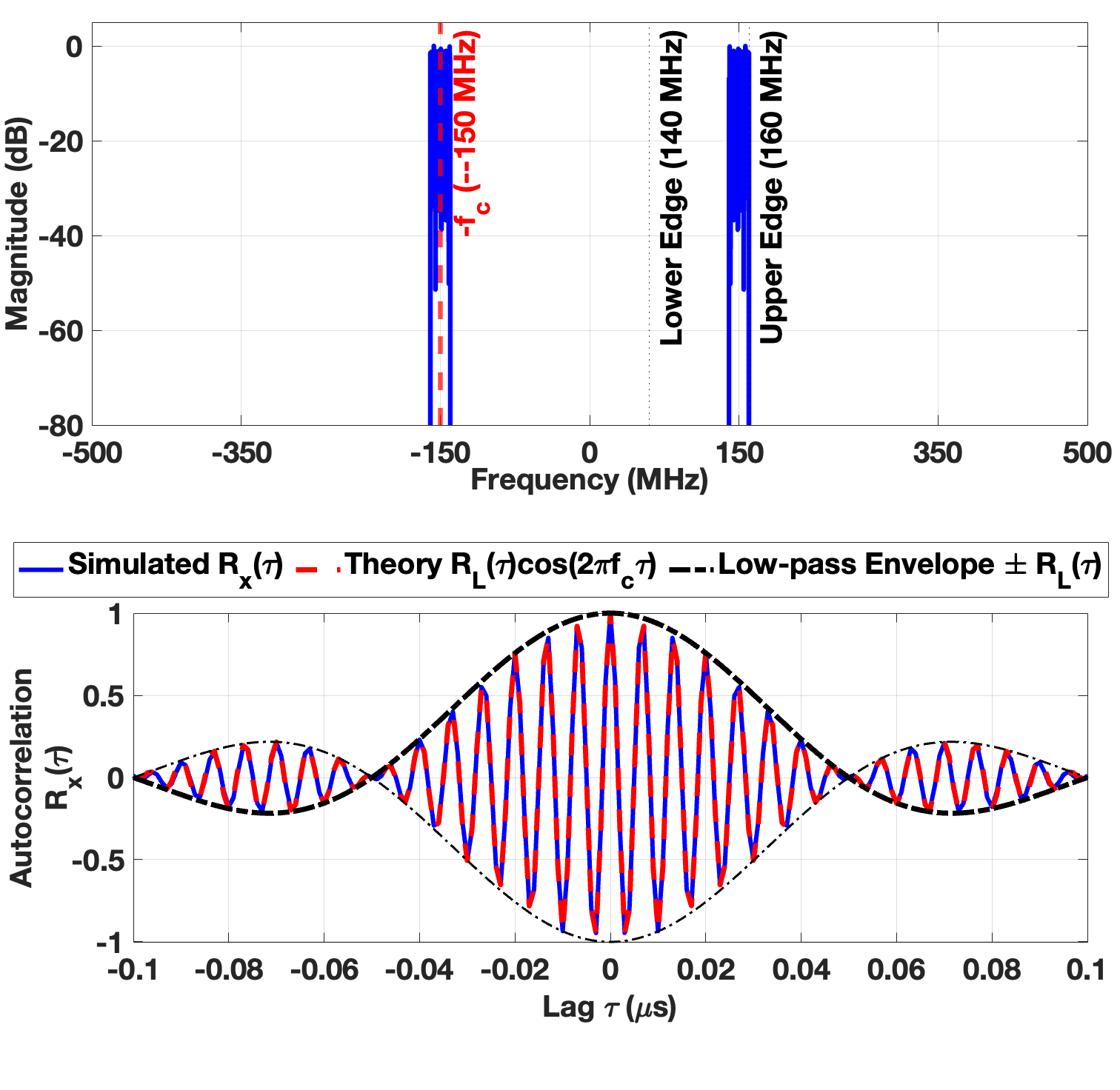} \vspace{-3em}
        \caption{}
    \end{subfigure}
    \hfill
    \begin{subfigure}[b]{0.48\textwidth}
        \centering
        \includegraphics[width=\textwidth]{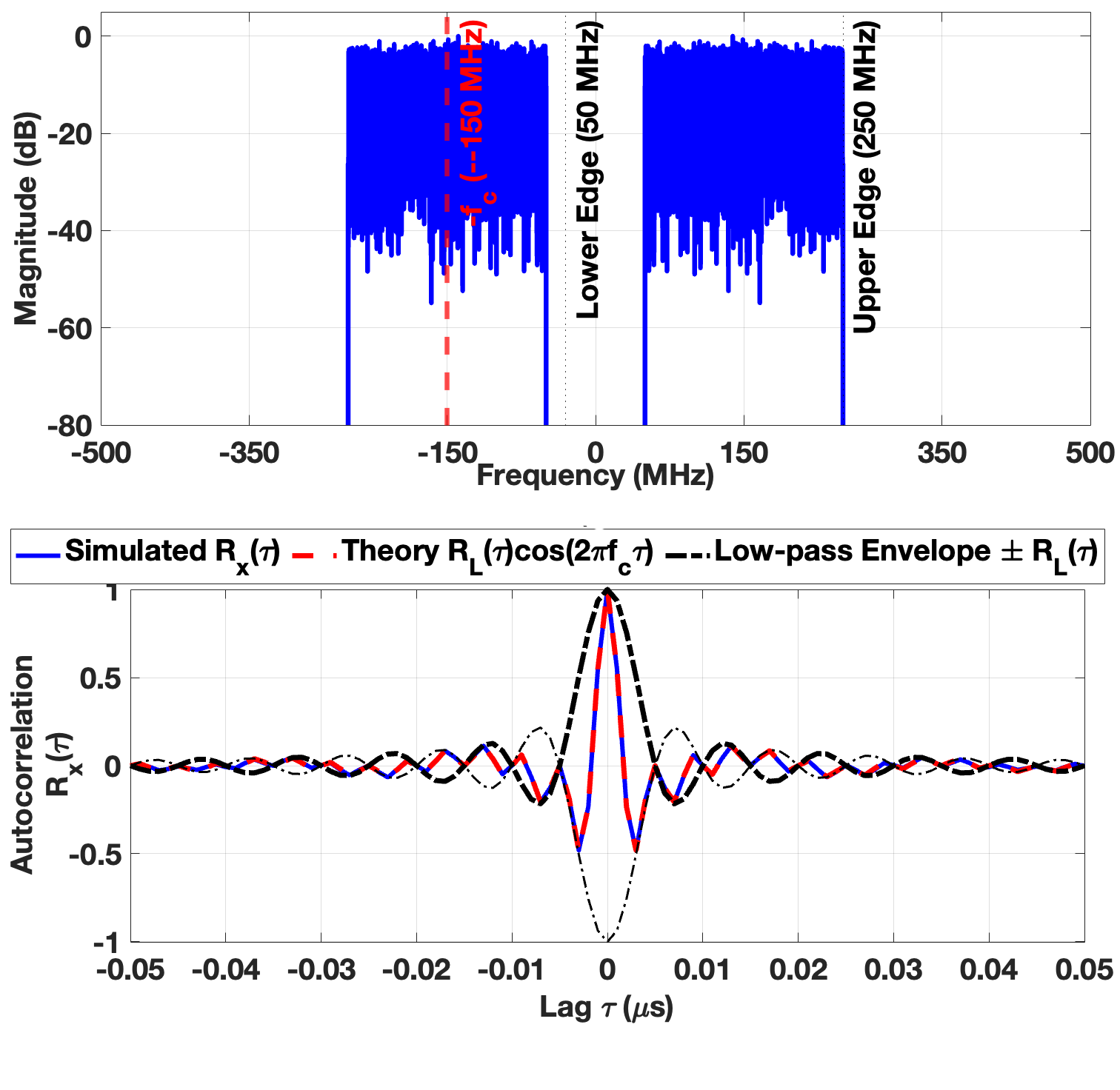} \vspace{-3em}
        \caption{}
    \end{subfigure}

    % Row 2
    \begin{subfigure}[b]{0.48\textwidth}
        \centering
        \includegraphics[width=\textwidth]{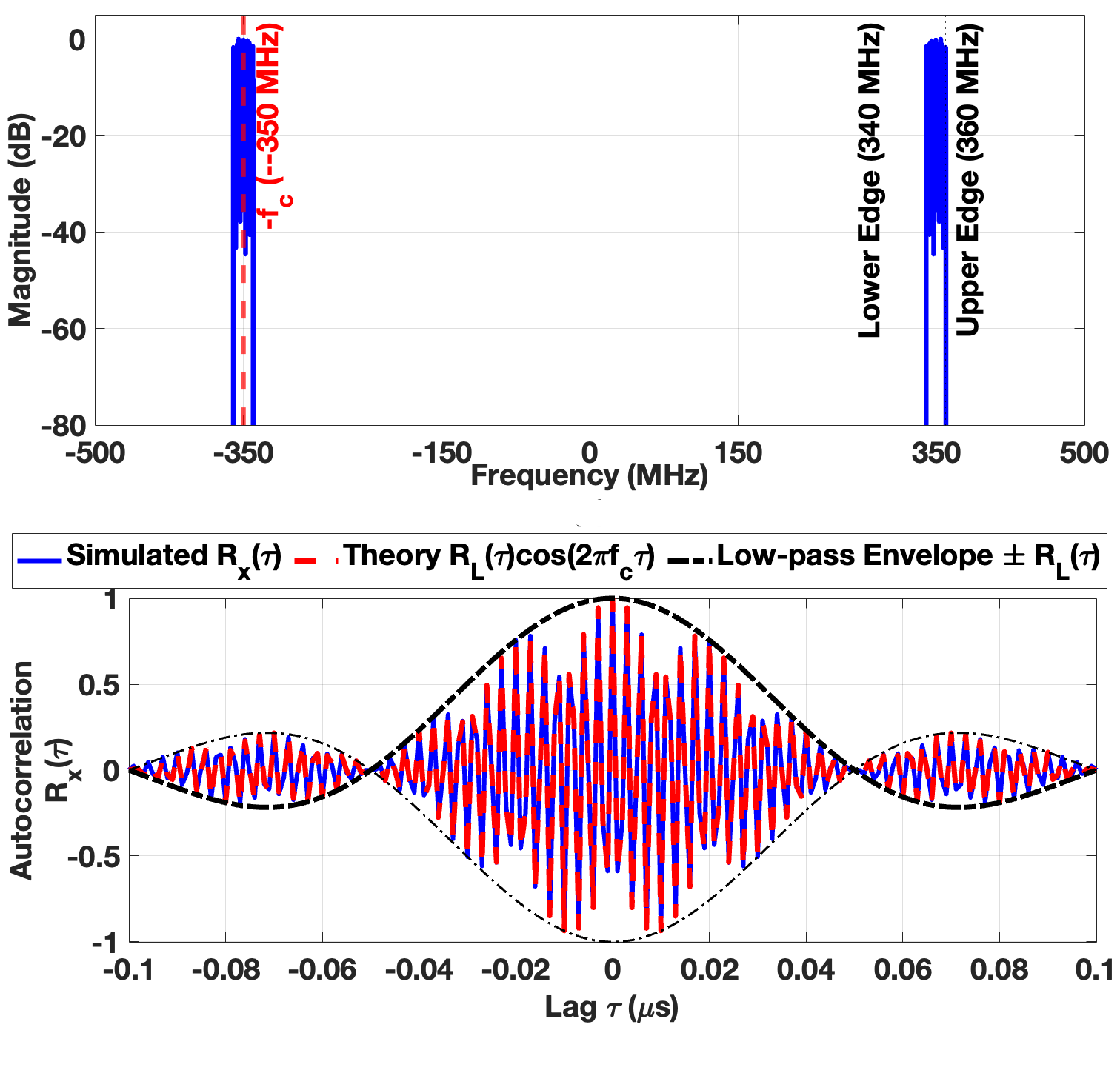} \vspace{-2.5em}
        \caption{}
    \end{subfigure}
    \hfill
    \begin{subfigure}[b]{0.48\textwidth}
        \centering
        \includegraphics[width=\textwidth]{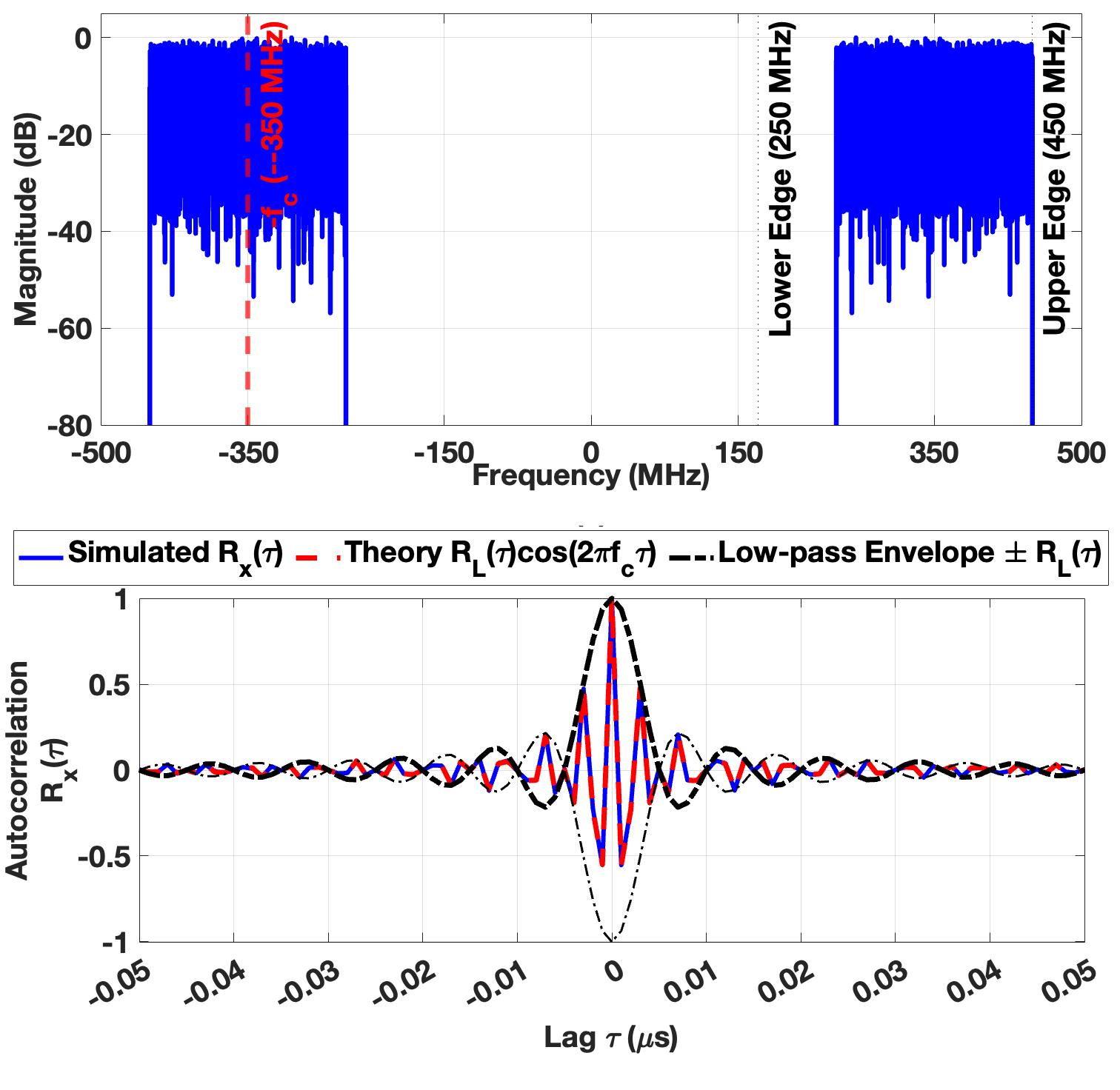} \vspace{-2.5em}
        \caption{}
    \end{subfigure}

    \captionsetup[sub]{labelformat=simple, labelsep=space}
    \renewcommand\thesubfigure{(\roman{subfigure})}
    \vspace{-0.5em}
    \caption{Two-sided spectrum and autocorrelation (simulated, theoretical, and low-pass envelope) across different lags of bandpass signals generated by filtering white noise and then upconverting them to carrier frequency, $f_c$ for: 
    (a) $f_c = 150$~MHz, bandwidth $= 10$~MHz; 
    (b) $f_c = 150$~MHz, bandwidth $= 100$~MHz; 
    (c) $f_c = 350$~MHz, bandwidth $= 10$~MHz; and 
    (d) $f_c = 350$~MHz, bandwidth $= 100$~MHz.} \vspace{-1em}
    \label{fig:SpectrumAutoCorrBandpass}
\end{figure*}

When the input signal is a bandpass signal centered at a high intermediate frequency $f_c$ with bandwidth $B$, its autocorrelation function modulates into a cosine-wave envelope\cite{peebles2015probability}:
\begin{equation}
   R_x(\tau) = R_L(\tau) \cos(2\pi f_c \tau)  
   \label{eq:wss_bandpass_autocorr}
\end{equation}
where $R_L(\tau)$ is the low-pass envelope ({\em e.g.}, a $\text{sinc}$ function if the bandpass spectrum is rectangular).\\
\indent To verify the autocorrelation expression, two types of wide-sense stationary (WSS) signals listed below were evaluated. 
\begin{enumerate}[label=(\roman*)]
    \item Bandpass-filtered white noise
    \item OFDM signals
\end{enumerate}
 Figure \ref{fig:SpectrumAutoCorrBandpass} presents the two-sided spectrum and autocorrelation for bandpass signals created by filtering white noise, sampling at $f_s = 1\text{ GHz}$, and upconverting to carrier frequencies of $f_c = 150\text{ MHz}$ and $350\text{ MHz}$ with one-sided bandwidths of $B = 10\text{ MHz}$ and $100\text{ MHz}$. As expected, the simulated autocorrelation $R_x(\tau)$ aligns exceptionally well with the theoretical $R_x(\tau)$ computed via \eqref{eq:wss_bandpass_autocorr} across these varying bandwidths and carrier frequencies. 

\begin{figure}
    %\vspace{-1em}
    \centering
    % Row 1
    \begin{subfigure}[b]{0.48\textwidth}
        \centering
        \vspace{-1em}
        \includegraphics[width=\textwidth]{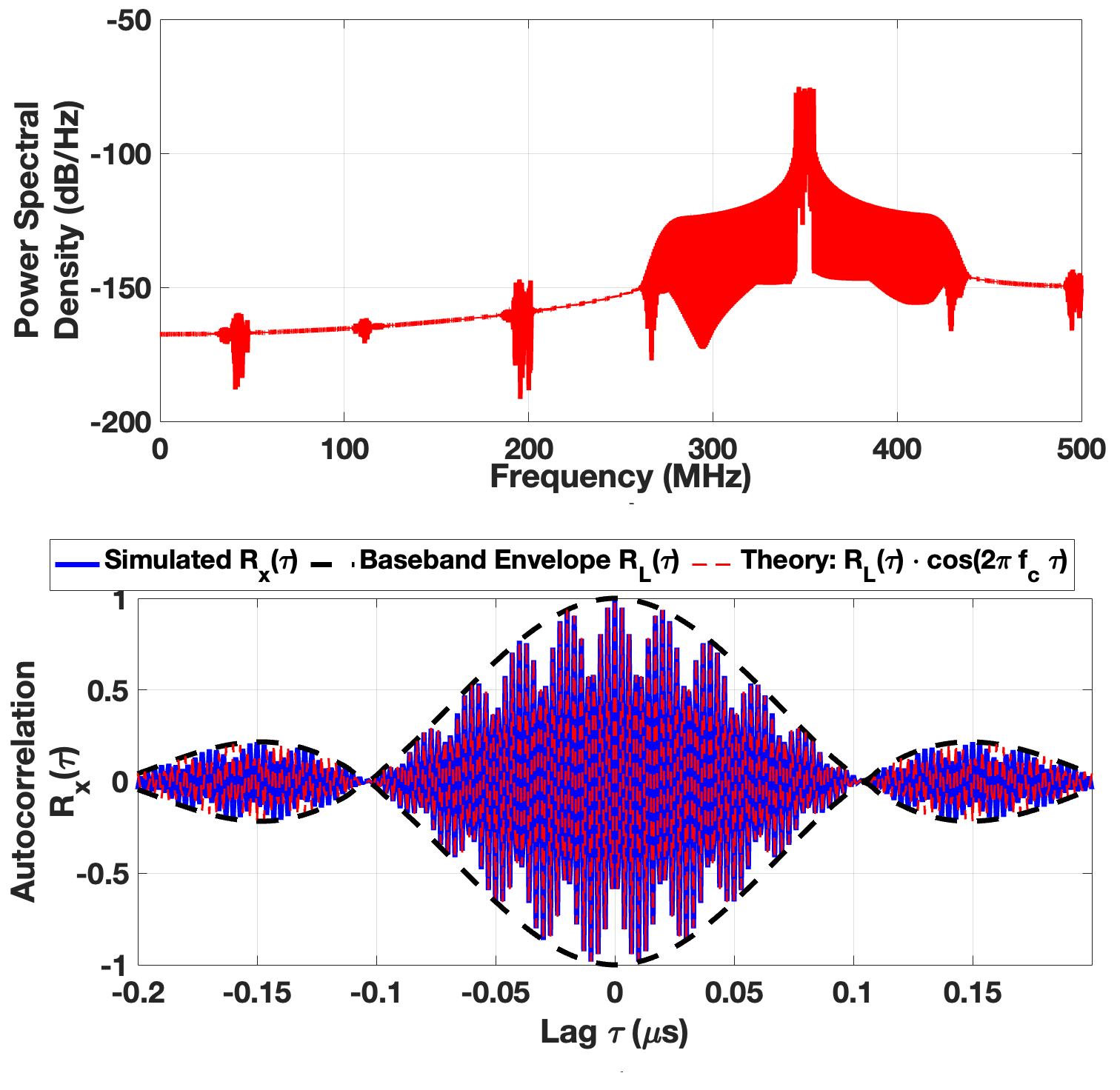} \vspace{-2.5em}
        \caption{}
    \end{subfigure}
    \hfill
    \begin{subfigure}[b]{0.48\textwidth}
        \centering
        \includegraphics[width=\textwidth]{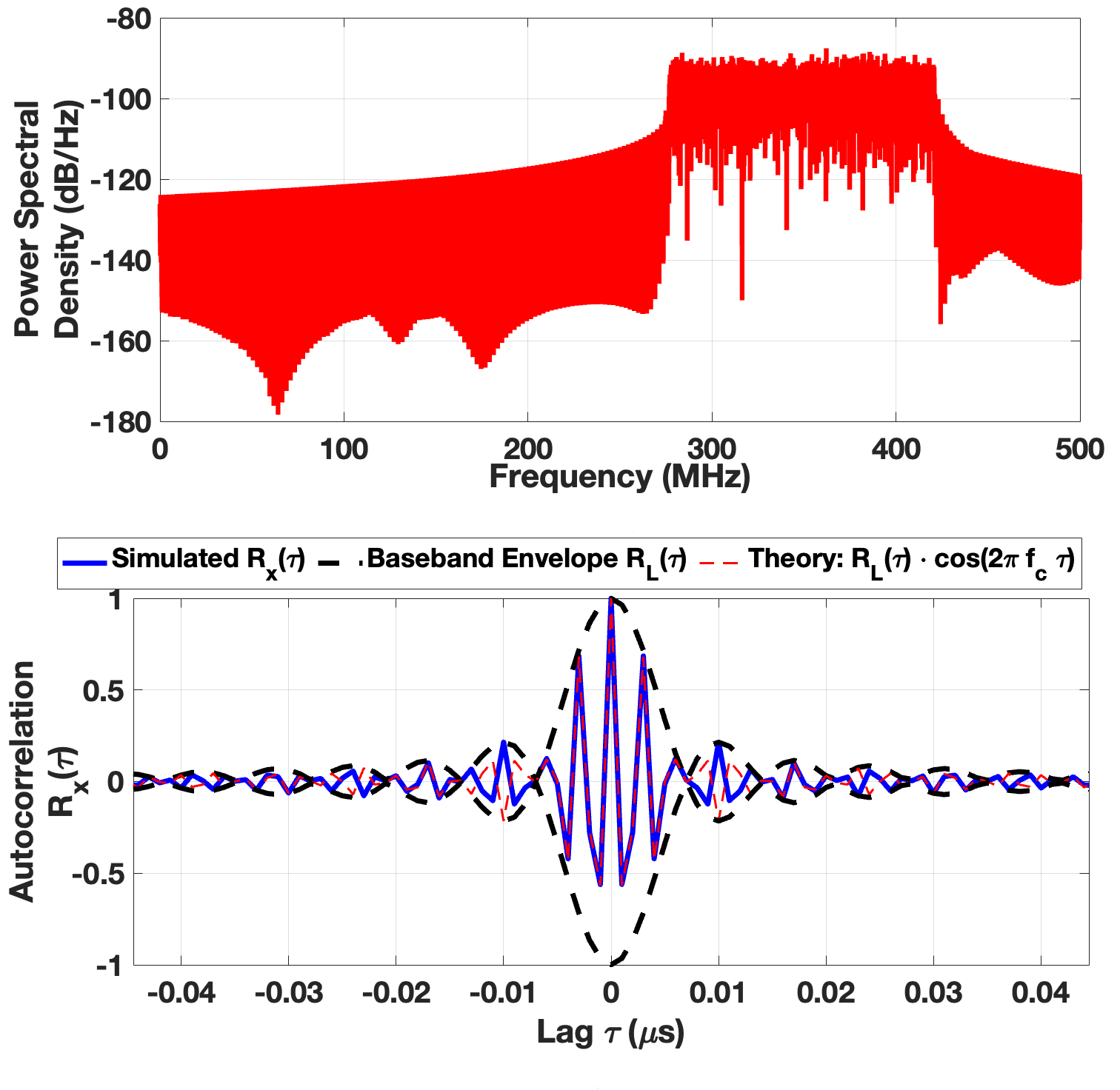} \vspace{-2.5em}
        \caption{}
    \end{subfigure}

    \captionsetup[sub]{labelformat=simple, labelsep=space}
    \renewcommand\thesubfigure{(\roman{subfigure})}
    \vspace{-0.5em}
    \caption{Two-sided spectrum and autocorrelation (simulated, theoretical, and low-pass envelope) across different lags of OFDM QPSK signals with 64-sub-carriers upconverted to carrier frequency, $f_c$ for: 
    (a) $f_c = 350$~MHz, bandwidth $= 9.6$~MHz; and 
    (b) $f_c = 350$~MHz, bandwidth $= 144$~MHz.} \vspace{-1.5em}
    \label{fig:SpectrumAutoCorrOfdm}
\end{figure}
Similarly, Figure \ref{fig:SpectrumAutoCorrOfdm} shows the spectrum and autocorrelation for 64-subcarrier OFDM QPSK signals at $f_c = 350\text{ MHz}$ with bandwidths of $B = 9.6\text{ MHz}$ and $144\text{ MHz}$ (corresponding to subcarrier spacings of $150\text{ kHz}$ and $2.25\text{ MHz}$, respectively). These results likewise demonstrate excellent agreement between the simulated and calculated autocorrelation values.

For high-frequency bandpass inputs, the dominant effect of a small timing skew $\Delta$ is a large phase shift in the carrier, while the change in the slow-moving envelope $R_L(\tau)$ is negligible ($R_L(T_s \pm \Delta) \approx R_L(T_s)$). Therefore, our primary cross-correlations can be written as:
\begin{equation}
\left.
\begin{aligned}
  R_{01} = R_x(T_s + \Delta) \approx R_L(T_s) \cos(2\pi f_c T_s + 2\pi f_c \Delta) \\ 
  R_{10} = R_x(T_s - \Delta) \approx R_L(T_s) \cos(2\pi f_c T_s - 2\pi f_c \Delta)  
\end{aligned}
\right\}
\label{eq:corr_wss_bps1}
\end{equation}
Before we proceed any further we need to obtain $R_L(T_S)$. In order to compute $R_L(T_S)$, let us construct the analytic signal of autocorrelation function $R_x(\tau)$ by taking its Hilbert Transform which is denoted by $\mathcal{H}\{R_x(\tau)\}$. Typically, the autocorrelation of a wideband/bandpass signal changes smoothly. As $R_L(\tau)$ is a slow-moving envelope relative to the high carrier frequency $f_c$ ({\em i.e.}, $B \ll f_c$), we can apply Bedrosian's Theorem \cite{bedrosian}, which says the Hilbert Transform of the baseband signal times the passband signal is the original baseband signal times the Hilbert transform of the passband signal. In this scenario, $R_L(\tau)$ is the baseband signal and $\cos(2\pi f_c T_s + 2\pi f_c \Delta)$ or $\cos(2\pi f_c T_s - 2\pi f_c \Delta)$ is the passband signal. Thus, we obtain the following: \vspace{-1em}
\begin{align}
    \mathcal{H}\{R_x(\tau)\} &= \mathcal{H}\{R_L(\tau) \cos(2\pi f_c \tau)\} \nonumber \\
    &\approx R_L(\tau) \sin(2\pi f_c \tau).
\end{align} 
Therefore, the complex analytic signal $\tilde{R}_x(\tau)$ is given as follows:
\begin{align}
    \tilde{R}_x(\tau) &= R_x(\tau) + j \mathcal{H}\{R_x(\tau)\} \nonumber \\
    &\approx  R_L(\tau) \left[ \cos(2\pi f_c \tau) + j \sin(2\pi f_c \tau) \right] \\
    &= R_L(\tau) e^{j 2\pi f_c \tau} \nonumber
\end{align} 
The magnitude of the analytic signal can be used to obtain $R_L(\tau)$ as shown below:
\begin{equation}
    R_L(\tau) = \vert{}\tilde{R}_x(\tau)\vert{} = \sqrt{R_x^2(\tau) + \left( \mathcal{H}\{R_x(\tau)\} \right)^2}
\end{equation}

In a TI-ADC, $R_{01}$ is not a continuous-time waveform, it is a single scalar correlation value evaluated at the interleaved lag $\tau = T_s + \Delta$, {\em i.e.}, $R_{01} = \mathbb{E}[x_0[n]x_1[n]] = R_x(T_s + \Delta)$. We can instead construct a vector of correlation values across multiple discrete delays $k \cdot T_s$, as given by the following equation,
\begin{equation}
    R_{01}[k] = E[x_0[n] \cdot x_1[n-k]], \quad k \in \{-L, \dots, L\}
\end{equation}
and then multiply this vector by a pre-computed set of odd-tap Hilbert filter coefficients given by the following equation: \vspace{-0.5em}
\begin{equation}
    \mathcal{H}\{R_{01}\} \approx \sum_{k \neq 0, \text{odd}} \frac{2}{\pi k} \cdot R_{01}[k]
    \label{eq:hilber_corr}
\end{equation}
As mentioned earlier, the autocorrelation of a wideband/bandpass signal changes smoothly. Thus, the Hilbert Transform in \eqref{eq:hilber_corr} can be computed with few filter coefficients. 

\indent Once, $\mathcal{H}\{R_{01}\}$ is computed, then $R_L(T_S)$ can be computed using the following equation:
\begin{equation}
    R_L(T_S) = \sqrt{R_{01} + \left( \mathcal{H}\{R_{01}\} \right)^2}
    \label{eq:rl_ts_computaion}
\end{equation}
Thus, we can rewrite \eqref{eq:corr_wss_bps1} as follows:
\begin{eqnarray}
    2\pi f_c T_s + 2\pi f_c \Delta = \overbrace{\cos^{-1}\left(\frac{R_{01}}{R_L(T_S)}\right )}^a \\
    2\pi f_c T_s - 2\pi f_c \Delta = \underbrace{\cos^{-1}\left( \frac{R_{12}}{R_L(T_S)} \right )}_b
\end{eqnarray}
By taking the sum and difference of the above two equations we get the following:
\begin{eqnarray}
    2\pi f_c T_s = \frac{a+b}{2}, \\
    2\pi f_c \Delta = \frac{a-b}{2}.
\end{eqnarray}
Dividing the above two equations we obtain,
%\begin{tcolorbox}[colback=gray!20, colframe=black]
\begin{equation}
    \frac{\Delta}{T_s} = \frac{a-b}{a+b}.
    \label{eq:wss_corr_closed_form}
\end{equation}
%\end{tcolorbox}

\begin{figure}[!h]
\centering
\resizebox{\linewidth}{!}{%
\begin{tikzpicture}[
  >=Stealth,
  node distance=0.7cm,
  thick,
  block/.style={
    draw,
    rectangle,
    rounded corners=3pt,
    line width=0.9pt,
    fill=blue!5,
    align=center,
    inner sep=6pt,
    font=\small,
    minimum width=6.8cm
  },
  startend/.style={
    draw,
    rectangle,
    rounded corners=10pt,
    line width=1pt,
    fill=gray!15,
    align=center,
    inner sep=6pt,
    font=\small\bfseries,
    minimum width=4.5cm
  },
  arrow/.style={
    ->,
    line width=0.9pt
  }
]

  % --- Flowchart Nodes ---
  \node[startend] (start) {Start / Input Signals $y_0[n], y_1[n]$};

  \node[block, below=of start] (step1) {
    \textbf{Compute Zero-Lag Cross-Correlations}\\[3pt]
    $R_{01}[0] = \mathbb{E}[y_0[n]y_1[n]]$\\[2pt]
    $R_{10}[0] = \mathbb{E}[y_1[n]y_0[n+1]]$
  };

  \node[block, below=of step1] (step2) {
    \textbf{Compute Odd-Lag Cross-Correlations}\\[3pt]
    $R_{01}[k] = \mathbb{E}[y_0[n]y_1[n-k]]$\\[2pt]
    for $k \in \{\pm 1, \pm 3, \pm 5, \pm 7, \pm 9\}$ \quad ($10$ terms)
  };

  \node[block, below=of step2] (step3) {
    \textbf{Approximate Hilbert Transform}\\[3pt]
    $\mathcal{H}\{R_{01}\} \approx \sum_{k \neq 0, \text{odd}} \frac{2}{\pi k} \cdot R_{01}[k]$
  };

  \node[block, below=of step3] (step4) {
    \textbf{Estimate Correlation Envelope}\\[3pt]
    $R_L(T_S) = \sqrt{R_{01}[0]^2 + \left( \mathcal{H}\{R_{01}\} \right)^2}$
  };

  \node[block, below=of step4] (step5) {
    \textbf{Compute Phase Angles}\\[3pt]
    $a = \cos^{-1}\left(\frac{R_{01}[0]}{R_L(T_S)}\right), \quad b = \cos^{-1}\left(\frac{R_{10}[0]}{R_L(T_S)}\right)$
  };

  \node[block, below=of step5] (step6) {
    \textbf{Compute Fractional Timing Offset}\\[3pt]
    $\frac{\Delta}{T_s} = \frac{a - b}{a + b}$
  };

  \node[startend, below=of step6] (end) {Output Timing Offset $\frac{\Delta}{T_s}$};

  % --- Connecting Arrows ---
  \draw[arrow] (start) -- (step1);
  \draw[arrow] (step1) -- (step2);
  \draw[arrow] (step2) -- (step3);
  \draw[arrow] (step3) -- (step4);
  \draw[arrow] (step4) -- (step5);
  \draw[arrow] (step5) -- (step6);
  \draw[arrow] (step6) -- (end);

\end{tikzpicture}%
}
\caption{Flowchart for computing the fractional timing offset $\Delta/T_s$.} \vspace{-1.5em}
\label{fig:timing_offset_flowchart}
\end{figure}
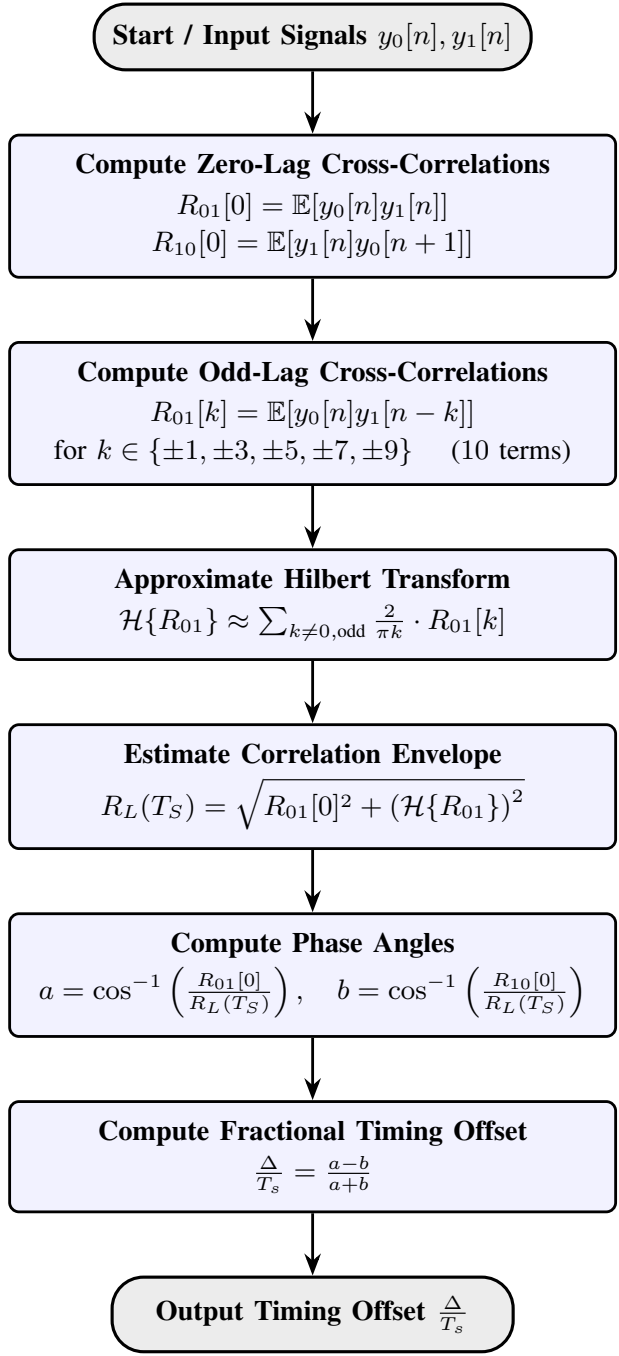
Fig. \ref{fig:timing_offset_flowchart} shows the flow-chart for computing the fractional timing offset $\Delta / T_S$ for WSS signals.

\begin{figure}[!h]
    \centering
    %\vspace{-2em}
    \includegraphics[width=0.48\textwidth]{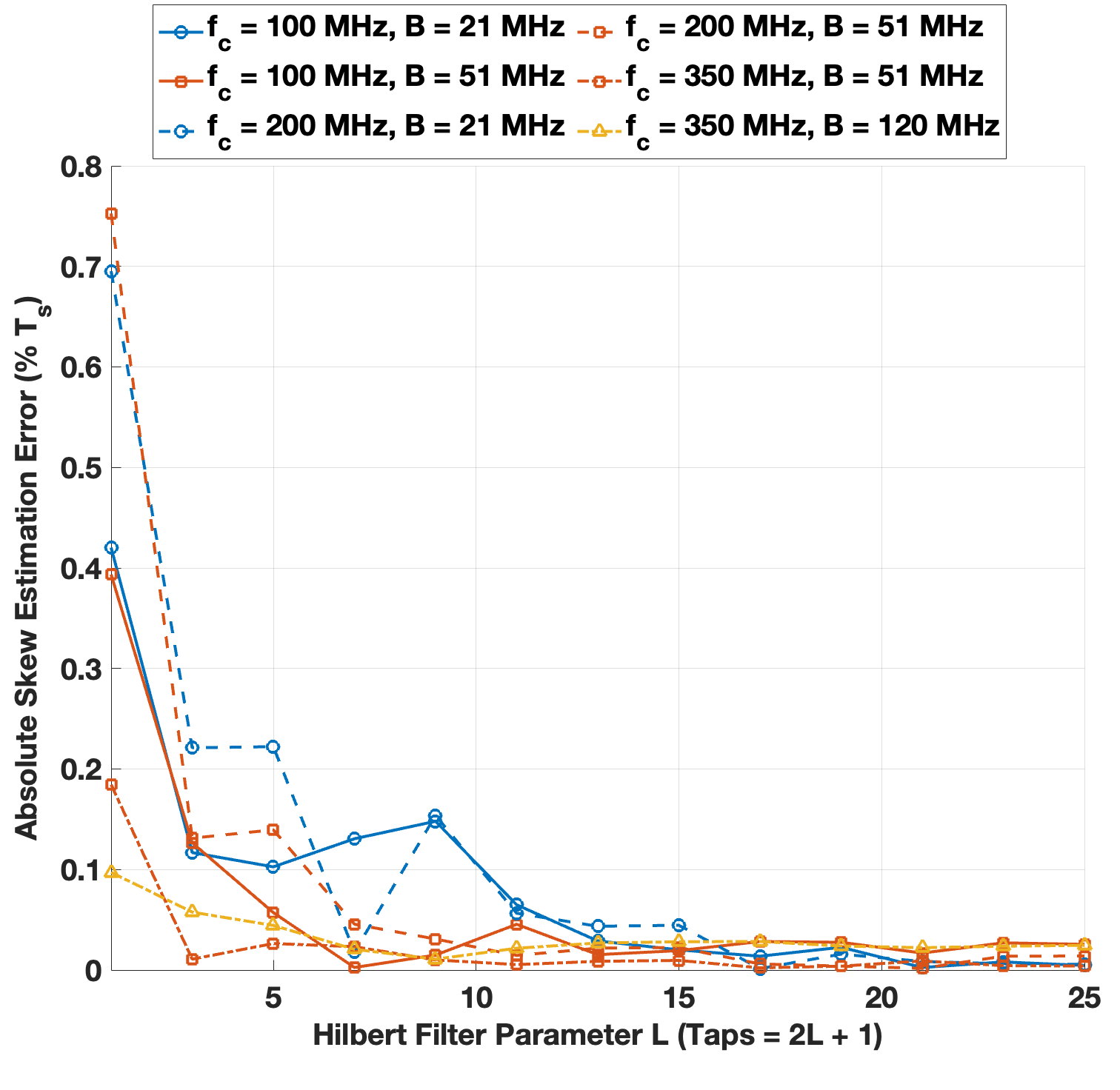} \vspace{-1em}
    \caption{Skew estimation error vs. Hilbert Transform filter parameter $L$ for OFDM signals of bandwidths $21$ MHz, $51$ MHz, and $120$ MHz and carrier frequency $f_c$ of 100 MHz, 200 MHz, and 350 MHz, sampled at $f_s=1$ GHz.} \vspace{-1.5em}
    \label{fig:hilbert}
\end{figure}

\indent Figure \ref{fig:hilbert} shows the skew estimation error vs. Hilbert Transform filter length ($L$) for OFDM signals of bandwidths $21$ MHz, $51$ MHz, and $120$ MHz and carrier frequency $f_c$ of 100 MHz, 200 MHz, and 350 MHz, sampled at $f_s=1$ GHz. Based on Fig. \ref{fig:hilbert} $L=11$ is optimal length of Hilber transform that has least error for signals up to $80\%$ of Nyquist.
To evaluate the proposed skew estimation algorithm (ref. Fig. \ref{fig:timing_offset_flowchart}), simulations were conducted using a two-channel TI-ADC model configured with the parameters listed below. The validation was performed across both previously defined WSS signal types: bandpass-filtered white noise and 64-subcarrier OFDM QPSK signals.
\begin{itemize}

    \item \textbf{System Sampling Rate:} $f_s = 1\text{ GSPS}$ ($T_s = 1\text{ ns}$)\item \textbf{Channel Sampling Rate:} $f_{\text{ch}} = f_s / 2 = 500\text{ MSPS}$
    
    \item \textbf{Quantization:} 12-bit ideal resolution, $V_{\text{ref}} = 1\text{ V}$ ($\pm 1\text{ V}$ full-scale range)
    
    \item \textbf{Timing Skew Range:} Swept from $0.55\%$ to $3.5\%$ of $T_s$ in increments of $0.3\%$, with absolute skew defined per \eqref{eq:per_skew}
    
    \item \textbf{Hilbert Filter Length:} $L = 11$\item \textbf{Autocorrelation Window:} $N = 16\,384$ samples ($2^{14}$)
    
    \item \textbf{Carrier Frequencies ($f\_c$):} $101\text{ MHz}$, $210\text{ MHz}$, $310\text{ MHz}$, and $395\text{ MHz}$
    
    \item \textbf{Signal Bandwidths ($B$):} $51\text{ MHz}$ and $101\text{ MHz}$
    
    \end{itemize}
    
Table~\ref{tab:skew_estimation_results} presents the estimated skew and corresponding residual-skew defined by \eqref{eq:res_skew} for the 64-subcarrier OFDM QPSK signals across various carrier frequencies, bandwidths, and input skew conditions.
\begin{align}
    \text{Residual Skew (\%)} =& \text{ Actual Skew (\%)} \nonumber \\ 
     &- \text{ Estimated Skew (\%)}
     \label{eq:res_skew}
\end{align}
\begin{table}[!h]
\caption{Two-channel TI-ADC Skew estimation using \eqref{eq:wss_corr_closed_form} for OFDM signals with different $f_c$ and bandwidth}
\begin{tabular}{|c|c|c|c|c|}
\hline
\textbf{\begin{tabular}[c]{@{}c@{}}Skew \\ \\ (\%)\end{tabular}} &
  \textbf{\begin{tabular}[c]{@{}c@{}}Carrier\\ Frequency\\ (MHz)\end{tabular}} &
  \textbf{\begin{tabular}[c]{@{}c@{}}Bandwidth\\ \\ (MHz)\end{tabular}} &
  \textbf{\begin{tabular}[c]{@{}c@{}}Estimated\\ Skew \\ (\%)\end{tabular}} &
  \textbf{\begin{tabular}[c]{@{}c@{}}Residual\\ Skew \\ (\%)\end{tabular}} \\ \hline
\multirow{6}{*}{0.550} & 101                  & 51  & 0.557 & -0.007 \\ \cline{2-5} 
                       & 210                  & 51  & 0.531 & 0.019  \\ \cline{2-5} 
                       & \multirow{2}{*}{310} & 51  & 0.556 & -0.006 \\ \cline{3-5} 
                       &                      & 101 & 0.540 & 0.010  \\ \cline{2-5} 
                       & \multirow{2}{*}{395} & 51  & 0.552 & -0.002 \\ \cline{3-5} 
                       &                      & 101 & 0.537 & 0.013  \\ \hline
\multirow{6}{*}{1.550} & 101                  & 51  & 1.594 & -0.044 \\ \cline{2-5} 
                       & 210                  & 51  & 1.498 & 0.052  \\ \cline{2-5} 
                       & \multirow{2}{*}{310} & 51  & 1.569 & -0.019 \\ \cline{3-5} 
                       &                      & 101 & 1.524 & 0.026  \\ \cline{2-5} 
                       & \multirow{2}{*}{395} & 51  & 1.555 & -0.005 \\ \cline{3-5} 
                       &                      & 101 & 1.515 & 0.035  \\ \hline
\multirow{6}{*}{2.450} & 101                  & 51  & 2.528 & -0.078 \\ \cline{2-5} 
                       & 210                  & 51  & 2.366 & 0.084  \\ \cline{2-5} 
                       & \multirow{2}{*}{310} & 51  & 2.479 & -0.029 \\ \cline{3-5} 
                       &                      & 101 & 2.409 & 0.041  \\ \cline{2-5} 
                       & \multirow{2}{*}{395} & 51  & 2.457 & -0.007 \\ \cline{3-5} 
                       &                      & 101 & 2.397 & 0.053  \\ \hline
\multirow{6}{*}{3.500} & 101                  & 51  & 3.617 & -0.117 \\ \cline{2-5} 
                       & 210                  & 51  & 3.380 & 0.120  \\ \cline{2-5} 
                       & \multirow{2}{*}{310} & 51  & 3.542 & -0.042 \\ \cline{3-5} 
                       &                      & 101 & 3.443 & 0.057  \\ \cline{2-5} 
                       & \multirow{2}{*}{395} & 51  & 3.509 & -0.009 \\ \cline{3-5} 
                       &                      & 101 & 3.428 & 0.072  \\ \hline
\end{tabular}
\label{tab:skew_estimation_results}
\end{table}
The following are some of the salient features of the proposed method outlined in the flow-chart given in Fig. \ref{fig:timing_offset_flowchart} that can be concluded from Table~\ref{tab:skew_estimation_results}:
\begin{itemize}
    \item \textbf{High Estimation Accuracy:} Across all evaluated input skews ($0.550\%$ to $3.500\%$), the estimated values track the actual timing skew closely. The residual skew remains strictly bounded within $-0.117\%$ to $+0.120\%$.
    
    \item \textbf{Carrier Frequency and Bandwidth Invariance:} The estimation error remains consistently minimal regardless of variations in carrier frequency ($f_c \in \{101, 210, 310, 395\}\text{ MHz}$) or signal bandwidth ($B \in \{51, 101\}\text{ MHz}$).
\end{itemize}
\subsection{Hardware Complexity Analysis}
\begin{figure}[!h]
    %\vspace{-1em}
    \centering
    % Row 1
    \begin{subfigure}[b]{0.48\textwidth}
        \centering
        \vspace{-1em}
        \includegraphics[width=\textwidth]{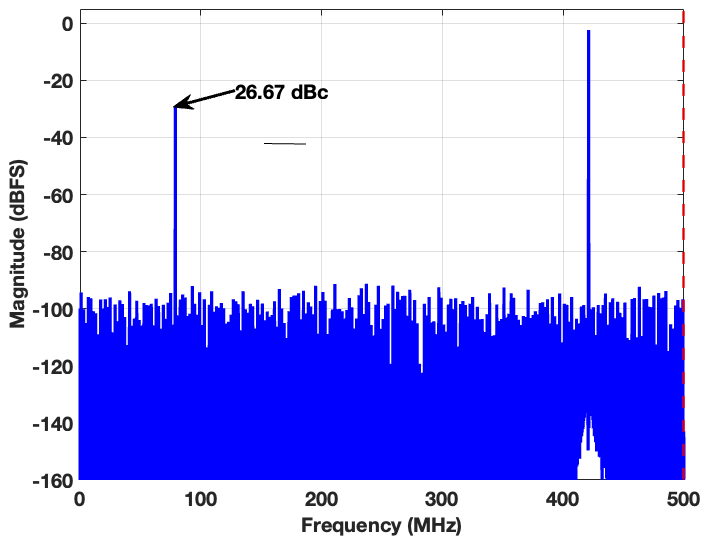}
        \caption{}
    \end{subfigure}
    \hfill
    \begin{subfigure}[b]{0.48\textwidth}
        \centering
        \includegraphics[width=\textwidth]{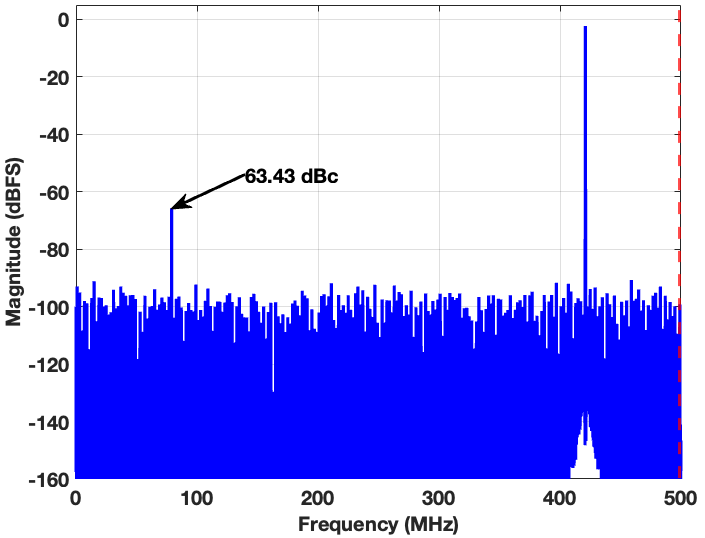}
        \caption{}
    \end{subfigure}
    \hfill
    \begin{subfigure}[b]{0.48\textwidth}
        \centering
        \includegraphics[width=\textwidth]{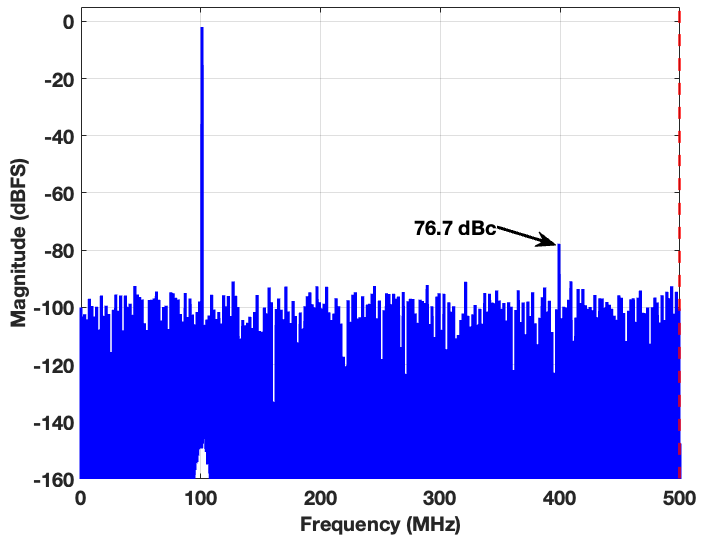}
        \caption{}
    \end{subfigure}
    \captionsetup[sub]{labelformat=simple, labelsep=space}
    \renewcommand\thesubfigure{(\roman{subfigure})}
    \vspace{-1.5em}
    \caption{Two-channel TI-ADC spectrum sine-wave (a) with $3.5$\% skew for $f_{in}=421$ MHz, (b) after skew correction with estimated skew of $3.45$\% for $f_{in}=421$ MHz, and (c) after skew correction with estimated skew of $3.45$\% for $f_{in}=101$ MHz.} \vspace{-1em}
    \label{fig:SkewEstSine}
\end{figure}

\begin{figure}[!h]
    \vspace{-0.5em}
    \centering
    % Row 1
    \begin{subfigure}[b]{0.48\textwidth}
        \centering
        \vspace{-0.5em}
        \includegraphics[width=\textwidth]{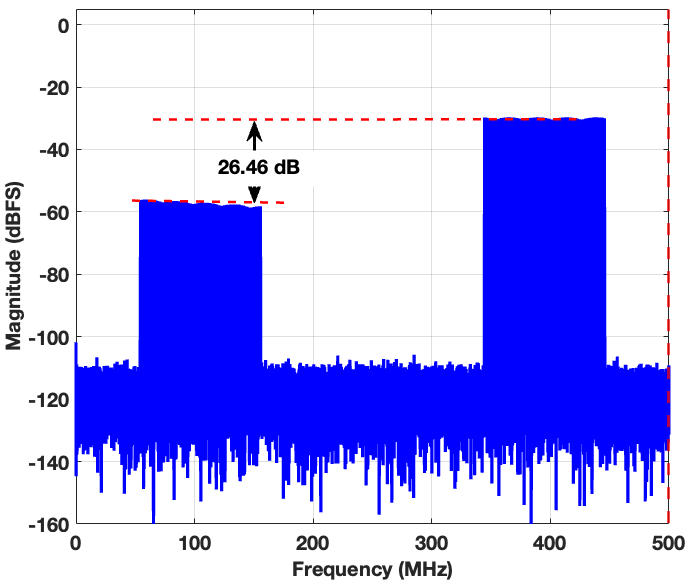}
        \caption{}
    \end{subfigure}
    \hfill
    \begin{subfigure}[b]{0.48\textwidth}
        \centering
        \includegraphics[width=\textwidth]{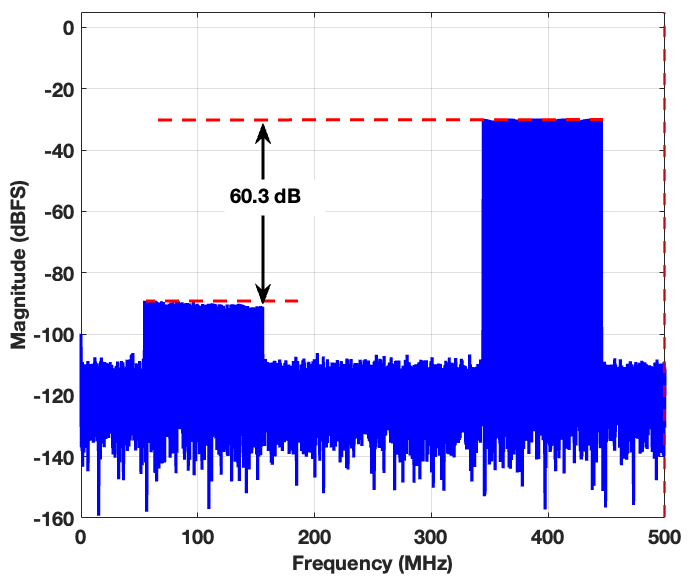}
        \caption{}
    \end{subfigure}

    \captionsetup[sub]{labelformat=simple, labelsep=space}
    \renewcommand\thesubfigure{(\roman{subfigure})}
    \vspace{-0.5em}
    \caption{Two-channel TI-ADC spectrum of OFDM QPSK signal (64-sub-carriers, $f_c=395$ MHz, and bandwidth $=101$ MHz) (a) with $3.5$\% skew and (b) after skew correction with estimated skew of $3.43$\%.} \vspace{-1em}
    \label{fig:SkewEstSpec1}
\end{figure}

For a two-channel TI-ADC ($M=2$) operating at a system sampling rate of $f_s = 1\text{ GSPS}$ ($f_{\text{ch}} = 500\text{ MSPS}$), with 12-bit ideal quantization and an autocorrelation averaging window of $N = 16\,384$ ($2^{14}$ samples), the proposed WSS skew estimation algorithm distributes its computational load into two distinct clock domains:

\begin{itemize}
    \item \textbf{High-Speed Operations ($f_{\text{ch}} = 500\text{ MSPS}$):}
    \begin{itemize}
        \item \textbf{Zero-Lag Cross-Correlations:} $M = 2$ Multiply-Accumulate (MAC) operations running continuously at $f_s / M = 500\text{ MHz}$.
        \item \textbf{Odd-Lag Cross-Correlations ($L = 11$):} 10 MAC operations for the odd lag terms $k \in \{\pm 1, \pm 3, \pm 5, \pm 7, \pm 9\}$, running continuously at $f_s / M = 500\text{ MHz}$.
    \end{itemize}

    \item \textbf{Low-Speed Operations (Decimated by $N = 16\,384$, running at $30.5\text{ kHz}$):}
    \begin{itemize}
        \item \textbf{Hilbert Transform Approximation:} A 10-tap FIR filter running at $f_s/(16384 \cdot M) \approx 30.5\text{ kHz}$.
        \item \textbf{Correlation Envelope Estimation:} A single square-root operation implemented using CORDIC at $30.5\text{ kHz}$.
        \item \textbf{Phase Angle Computation:} Two 256-row\footnote{The 256 row LUT is based on the fact that the estimated skew would control a Digital Controlled Delay Line whose resolution is $8$-bits \cite{HseihMediatek2025}} Lookup Tables (LUTs) for evaluating $\cos^{-1}(\cdot)$ operations at $30.5\text{ kHz}$.
        \item \textbf{Fractional Timing Offset Calculation:} A single division evaluated using a 6-iteration Newton-Raphson (NR) \cite{Tjalling1995} algorithm at $30.5\text{ kHz}$.
    \end{itemize}
\end{itemize}
\begin{table*}[!t]
\centering
\caption{Comparison of Hardware Complexity with other all digital skew-estimation} \vspace{-0.5em}
\label{tab:hardware_complexity}
\begin{tabular}{|p{1.8cm}|p{6.9cm}|p{6.5cm}|}
\hline \hline
\textbf{Architecture Aspect} & \textbf{Proposed Algorithm} & \textbf{Derivative-Based Algorithm (ref. \eqref{eq:delta_wss})} \cite{luu2018design,LeDortz2014,Divi2009} \\ \hline \hline
\textbf{High-Speed Computation ($f_{\text{ch}}$)} & 
Requires 12 MAC operations (2 zero-lag and 10 odd-lag terms). & 
Requires $\approx 31$ MAC operations (2 zero-lag MACs, 28 MACs to realize 28-tap FIR filter to compute the derivative using Parks-McClellan method, and 1 MAC to compute $R_x'(T_s) = \mathbb{E}[y_0[n]y_1'[n]]$ (ref. \eqref{eq:auto_deriv}). \\ \hline
\textbf{Filter Constraints} & 
Eliminates high-speed differentiation; utilizes a lightweight 10-tap FIR filter operating at a heavily decimated rate ($f_{ch}/N$, where $N$ is the window size of correlation). & 
Requires a high-order, wideband derivative FIR filter operating at full channel rate ($f_{ch}$), increasing power consumption and bandwidth limitations. \\ \hline
\textbf{Low-Speed Post-Processing} & 
Subsampled by $N$ the correlation window size ; uses low-cost CORDIC, two 256-row LUTs., and a 6-iteration NR-divider. & 
Subsampled by $N$; requires a low-speed divider to compute $\Delta \approx \frac{R_{01} - R_{12}}{2 R'_x(T_s)}$ (ref. \eqref{eq:delta_wss}) based on 6-iterations NR-method. \\ \hline \hline 
\end{tabular}
\vspace{-1.5em}
\end{table*} 
Figure \ref{fig:SkewEstSine} displays the spectrum of a 2-channel TI-ADC digitizing sine wave of $421$ MHz and $101$ MHz both before and after skew correction. With an initial $3.5\%$ timing skew, the image power is $26.67\text{ dBc}$ for $421$ MHz sine-wave. Applying the estimated skew correction ($3.45\%$) suppresses the image to $63.43\text{ dBc}$ and $76.7$ dBc for $421$ MHz and $101$ MHz sine-wave respectively, yielding an improvement of $\approx 37\text{ dB}$ and $\approx 50\text{ dB}$, respectively. \\
\indent Figure \ref{fig:SkewEstSpec1} displays the spectrum of a 2-channel TI-ADC digitizing an OFDM-QPSK signal (64 subcarriers, $f_c = 395\text{ MHz}$, bandwidth $= 101\text{ MHz}$) both before and after skew correction.
With an initial $3.5\%$ timing skew, the image power lies $26.45\text{ dB}$ below the main signal spectrum. Applying the estimated skew correction ($3.43\%$) suppresses the image to $60.3\text{ dB}$ below the main signal, yielding an improvement of $\approx 34\text{ dB}$.

\indent \indent Table \ref{tab:hardware_complexity} compares the hardware complexity of the proposed approach with the digital-skew estimation based on computing the derivative of the signal \cite{luu2018design,Dortz2014,Divi2009}, highlighting a clear trade-off between high-speed front-end requirements and low-speed back-end processing for an all digital skew-estimation. The proposed algorithm eliminates the power-intensive, high-order derivative FIR filter operating at the channel rate of $f_{ch}$-a key bottleneck in traditional derivative-based estimation given by \eqref{eq:delta_wss} and replaces it with $12$ parallel MAC operations for zero-lag and odd-lag cross-correlations. Although the proposed method introduces additional back-end arithmetic stages, including a 10-tap FIR filter, CORDIC square-root computation, two 256-row LUTs, and a Newton-Raphson divider, these operations are heavily decimated by the correlation window length, $N$ and run at a negligible clock rate of $f_{ch}/N$. Consequently, by shifting computational complexity from the high-speed sample rate to a low-speed digital domain, the proposed approach significantly reduces overall dynamic power consumption and area overhead while removing input bandwidth restrictions. 
\section{Conclusion}
\label{sec:conclusion}
This paper presents a low-complexity all digital background timing-skew estimation technique for two-channel TI-ADCs that operates effectively across diverse signal profiles without requiring resource-heavy derivative filters. By leveraging adjacent-channel cross-correlation alongside single-channel autocorrelation, closed-form analytical expressions were derived for both sinusoidal and Wide-Sense Stationary (WSS) inputs. 

The analytical models were validated through extensive simulations using bandpass-filtered noise and 64-subcarrier OFDM QPSK signals. For WSS inputs, the proposed algorithm demonstrated exceptional accuracy, maintaining residual skew errors within a narrow bound of $-0.117\%$ to $+0.120\%$ across input skews ranging from $0.550\%$ to $3.500\%$ of the sampling period. Furthermore, the estimation performance proved highly robust and invariant to changes in carrier frequency ($101\text{--}395\text{ MHz}$) and signal bandwidth ($51\text{--}101\text{ MHz}$). Lastly, an initial exploration of the implementation challenges and advantages of the proposed approach has been discussed. 

These results confirm that the proposed technique provides a computationally efficient and reliable framework for real-time background calibration in high-speed, multi-gigasample TI-ADC systems.
\vspace{-1em}
\section*{Acknowledgement}
The authors would like to acknowledge the use of Google-Gemini to facilitate MATLAB coding for the verification of the proposed methodology. \vspace{-1em}

\bibliographystyle{IEEEtran}
\bibliography{references}
\end{document}